\documentclass[usegraphicx,usenatbib,useAMS]{mn2e}\usepackage{times}
\usepackage{graphicx}
\usepackage{amssymb}
\usepackage{caption}
\let\url\relax

\def\apj{{ApJ}}
\def\apjs{{ApJs}} 
\def\apjl{{ApJL}}
\def\aap{{A.\&A}}
\def\aj{{AJ}}

\def\MNRAS{{MNRAS}}

\def\mnras{{MNRAS}}
\newcommand{\be}{\begin{equation}}
\newcommand{\ba}{\begin{eqnarray}}
\newcommand{\ee}{\end{equation}}
\newcommand{\ea}{\end{eqnarray}}

\begin{document}

\title [The halo mass function through the cosmic ages]{The halo mass function through the cosmic ages}
 
\author[W. A. Watson et al.]{William~A.~Watson$^1$\thanks{e-mail: 
W.Watson@sussex.ac.uk}, Ilian~T.~Iliev$^1$, Anson~D'Aloisio$^2$, Alexander~Knebe$^3$, \newauthor Paul~R.~Shapiro$^2$ and Gustavo~Yepes$^3$
\\
$^1$ Astronomy Centre, Department of Physics \& Astronomy, Pevensey II 
Building, University of Sussex, Falmer, Brighton, BN1 9QH, United Kingdom\\
$^2$ Department of Astronomy and Texas Cosmology Center, University of Texas, 
Austin, TX 78712, USA\\
$^3$ Departamento de F\'isica Te\'orica, Modulo C-XI, Facultad de Ciencias, 
Universidad Aut\'onoma de Madrid, 28049 Cantoblanco, Madrid, Spain
}

\date{\today} \pubyear{2012} \volume{000}
\pagerange{1} \twocolumn \maketitle
\label{firstpage}

\begin{abstract}

In this paper we investigate how the dark matter halo mass function evolves with redshift, based on a suite of very large (with $N_p = 3072^3 - 6000^3$ particles) cosmological N-body simulations. Our halo catalogue data spans a redshift range of $z = 0-30$, allowing us to probe the mass function from the Dark Ages to the present. We utilise both the Friends-of-Friends (FOF) and Spherical Overdensity (SO) halofinding methods to directly compare the mass function derived using these commonly used halo definitions. The mass function from SO haloes exhibits a clear evolution with redshift, especially during the recent era of dark energy dominance ($z < 1$). We provide a redshift-parameterised fit for the SO mass function valid for the entire redshift range to within $\sim 20\%$ as well as a scheme to calculate the mass function for haloes with arbitrary overdensities. The FOF mass function displays a weaker evolution with redshift. We provide a `universal' fit for the FOF mass function, fitted to data across the entire redshift range simultaneously, and observe redshift evolution in our data versus this fit. The relative evolution of the mass functions derived via the two methods is compared. For an SO halo defined via an overdensity of 178 versus the background matter density and an FOF halo defined via a linking length of 0.2 times the mean inter-particle separation we find that the mass functions most closely match at $z=0$. The disparity at $z=0$ between the FOF and SO mass functions resides in their high mass tails where the collapsed fraction of mass in SO haloes is $\sim80\%$ of that in FOF haloes. This difference grows with redshift so that, by $z>20$, the SO algorithm finds a $\sim50-80\%$ lower collapsed fraction in high mass haloes than the FOF algorithm.
\end{abstract}


\section{Introduction}
\label{intro:sect}

The abundance of dark matter haloes -- the halo mass function -- plays an important role in cosmology due to its sensitivity to a number of important parameters including the matter density of the universe, $\Omega_m$, the Hubble parameter, $h$, the spectral index of the primordial power spectrum, $n_s$, and the dark energy equation of state \citep[e.g.][]{Holder:2001db,Haiman:2001df,Weller:2001gk}. The mass function is currently difficult to pin down with much precision observationally (for recent results see \cite{2008ApJ...679L...1R,2009ApJ...692.1060V,2010ApJ...708..645R} and for an overview of the observational challenges see \cite{Eke:2005za}). The two main issues faced by observers in measuring the mass function are 1) building large enough datasets of galaxies or clusters to reduce statistical uncertainties that arise from low number counts, 2) relating the distribution of observed light to the underlying distribution of mass. In the case of galaxies the first problem is becoming much less of an issue as we perform larger and larger surveys \citep{2011MNRAS.413..971D,2011ApJ...741....8C,2012arXiv1211.0310L,2013AJ....145...10D}. The latter problem is overcome via studying the bias of galaxies as a tracer of the underlying matter peaks, a topic that has been well studied (see \cite{2011ApJ...741...19M} for a discussion of how accurate our current understanding is). For the case of clusters, both of the above observational issues represent major challenges. The number of observed clusters remains low despite recent, ongoing and future surveys \citep{2009A&A...497..667V,2009ApJ...692.1060V,2010ApJ...708..645R,2011MNRAS.413..386B, 2011A&A...534A.120T,2011AJ....141...94G,2012MNRAS.422...44P,2012MNRAS.423.1024M} and cluster counting via the Sunyaev-Zel'dovich (SZ) effect \citep{2010ApJ...722.1148F,2011A&A...536A...8P,2013arXiv1301.0816H,2013arXiv1303.5089P}. Relating cluster observables to their underlying masses is a well-studied problem in cosmology \citep{2009ApJ...692.1060V,2011A&A...536A..11P,2011A&A...536A..10P,2012ApJ...760...67R,2012arXiv1204.6305R,Angulo:2012ep,2013arXiv1303.5089P}. The existence of an intrinsic scatter in the relationships between observables and masses highlights that trying to obtain an accurate measurement of the high-mass tail of the mass function is a stiff observational challenge. 

In contrast, the mass function is relatively easy to probe via N-body cosmological simulations (see \cite{Lukic:2007fc} for a review of older work, more recent studies include \cite{2008MNRAS.385.2025C,Tinker:2008ff, Crocce:2009mg, Courtin:2010gx, Bhattacharya:2010wy, Angulo:2012ep}) and can, to an extent, be understood and modelled through analytic arguments \citep{1974ApJ...187..425P, 1991ApJ...379..440B,Lee:1998mh,2002MNRAS.329...61S,Maggiore:2009rv, Corasaniti:2011dr,Lim:2012jc}. The halo mass function -- derived from either analytic arguments or simulations -- is employed widely in astronomy: at low redshifts it is used in statistical analyses of cluster surveys to constrain cosmological parameters \citep{Smith:2011vm} as these large objects probe the high mass end of the function (for a recent review paper on the formation of galaxy clusters, including the role of the mass function, see \cite{Kravtsov:2012zs}). It is a key component in studies employing correlation functions between galaxies, as the halo-halo term directly depends on the mass function (see for example \cite{Cooray:2002dia} and references therein). At higher redshifts it is applied in modelling the process of reionization, which proceeded between $z\sim6-20$ from sources residing in dark matter haloes including population-III stars, early galaxies and accreting black holes. Any significant observed deviation from the mass function predicted by the $\Lambda$CDM cosmological model would create tension in our current understanding of structure formation in the universe. For example we are able to put bounds on how massive the largest observable clusters in the visible universe should be from the mass function \citep[see][and references therein]{Harrison:2011ep}, the discovery of larger clusters would require explanation.

Much has been been written on the topic of the mass function by groups working with N-body simulations. To date the majority of this work has looked at the mass function at low redshift ($z \lesssim 2$) using haloes derived using the Friends-of-Friends (FOF) algorithm \citep{1985ApJ...292..371D}. Despite this there has been a significant amount of work investigating the mass function at high redshift \citep{2006ApJ...642L..85H,Reed:2006rw,Lukic:2007fc,2008MNRAS.385.2025C,2011ApJ...740..102K} and various authors have studied the mass function derived via the alternative Spherical Overdensity (SO) algorithm \citep{1994MNRAS.271..676L}. For example \cite{Tinker:2008ff} calculate a fitted mass function for SO haloes valid for $z\lesssim2$; \cite{2001MNRAS.321..372J} employed both the FOF and SO algorithms finding similar results for each for $z<5$ (although the SO results had more scatter across redshifts and cosmologies, see the discussion on universality below); \cite{2008MNRAS.385.2025C} found good agreement between the FOF and SO mass functions at $z=10$ for appropriately chosen algorithm parameters (see below and \S~\ref{LL_and_Delta:sect}); \cite{Reed:2006rw} focussed on FOF haloes but compared the results to SO haloes finding the number density of haloes to be systematically lower in the case of SO haloes. Fitted functions are abundant in the literature (see \S~\ref{sect:comp}) with the majority being calibrated using FOF haloes. There is currently no robust mass function available for haloes derived using the SO algorithm at redshifts higher than $z\sim2$ which represents a gap in the literature that we address in this work. We aim in this paper to add our own fits for both FOF and SO halo mass functions and to exhibit their differences across a range of masses and redshifts.

There are shortcomings of both halofinding algorithms that the reader should be aware of. This is reflective of the fact that there is no specific definition of a dark matter halo that is agreed upon in the literature. The FOF algorithm identifies regions that are bounded by constant density contours in real space (see \S~\ref{finder_comparison:sect}) whereas the SO algorithm creates an artificial, spherically-bounded region; both of these outputs are referred to as `haloes'. In reality a dark matter halo is never perfectly spherical, and at higher masses and redshifts the virialisation process that occurs in high-density peaks is often incomplete, creating a tension between objects identified via the SO algorithm and physical reality. In the FOF case there is a systematic effect inherent to the algorithm known as `overlinking' which occurs when two haloes are linked together by a bridge of particles. When this occurs the resultant objects are not physically interpretable as virialised collapsed density peaks, but rather they are a representation of a complex system that is undergoing relaxation or merging. Previous authors \citep{1985ApJ...292..371D,1995ApJ...455....7M,1996MNRAS.281..716C,Lukic:2008ds} have calculated that 15-20\% of all FOF haloes calculated with the standard linking length of 0.2 are objects that have been bridged together in this manner. However it is important to note that for a suitable choice of linking length it is possible to obviate the effect of overlinking: in general the lower the linking length parameter the lower the amount of overlinking. The correspondence between SO haloes and FOF haloes has been studied, for example in \cite{2001A&A...367...27W} and \cite{2002ApJS..143..241W}, and recently empirical relations have been calculated by \cite{Lukic:2008ds}, \cite{Courtin:2010gx} and \cite{More:2011dc} that relate SO and FOF haloes in terms of their masses. This is looked at in more detail in \S~\ref{finder_comparison:sect} where we discuss the differences between the two halo types and the effect this has on the mass functions derived from them. In this paper we make no attempt to quantify the various underlying effects that lead to different mass function results when derived via the two algorithms. The effects, outlined in \S~\ref{hm_redshift_dep} and discussed in detail in \S~\ref{sect:halofindercomp}, are numerous and their interplay complex.

\cite{2001MNRAS.321..372J} published results suggesting that the mass function was perhaps `universal', i.e. independent of redshift or cosmology, when expressed in suitable units. A careful study by \cite{2002ApJS..143..241W} (building on earlier results on defining halo mass \citep{2001A&A...367...27W}) showed that deviations from a universal form were small but existent. Other negative results have been published by various authors \citep{Reed:2006rw,Tinker:2008ff, Crocce:2009mg, Courtin:2010gx, Bhattacharya:2010wy}. It was originally noted by \cite{2001MNRAS.321..372J} that in order to produce a universal mass function from FOF haloes one needs to take a constant linking length -- the key parameter that the FOF algorithm uses, see  \S~\ref{halofining:sect} -- across all redshifts or cosmologies being considered. \cite{Tinker:2008ff} observed a clearer departure from universality for SO haloes than for FOF haloes (in line with the earlier result of \cite{2001MNRAS.321..372J}), a fact that \cite{More:2011dc} propose is due to the effect of taking a fixed overdensity criterion -- the key parameter of the SO algorithm, see  \S~\ref{halofining:sect} -- rather than specific overdensity criteria bespoke to a given redshift or cosmology. One is therefore left with a choice: either adjust one's halo definitions to attempt to produce a universal mass function or keep one's halo definitions fixed and expect the function to deviate from a universal form. In this work we choose the latter. Deciding upon a suitable `universal' redshift-dependent halo overdensity or linking length parameter is still an open question (although \cite{Courtin:2010gx} and \cite{More:2011dc} go some way towards providing a solution), due to complicating factors such as the triaxiality of haloes in the SO case -- a purely spherical halo is an ideal that is never realised -- and overlinking in the FOF case. We shy away from demanding a universal mass function and pin our halo definitions down firmly in \S~\ref{halofining:sect}.

This paper is laid out as follows. In \S~\ref{sim:sect} we describe the simulations we have used to construct the mass function. In \S~\ref{halofining:sect} we outline the halo-finding methods we employed and the definitions of what we are referring to by the term `halo'. In \S~\ref{hmf:sect} we present our mass function results and fits. In \S~\ref{summary:sect} we present a summary of the work and a discussion of various points that arise.

\section{Simulations}
\label{sim:sect}

The simulations undertaken in this work are summarised in Table~\ref{summary_N-body_table}. They follow between 3072$^3$ (29 billion) to 6000$^3$ (216 billion) particles in a wide range of box sizes from 11.4~$h^{-1}$Mpc up to 6~$h^{-1}$Gpc. Spatial resolutions range from 0.18~$h^{-1}$kpc to 50~$h^{-1}$kpc while particle masses range from $3.6\times10^{3}h^{-1}\mathrm{M}_\odot$ to $7.5\times10^{10}h^{-1}\mathrm{M}_\odot$. This allows dark matter haloes to be resolved in a very large mass range, with a low end of $7.3\times10^{4}h^{-1}\mathrm{M}_\odot$ (for a 20 particle halo) in our smallest volumes, and no upper limit in the halo mass, since our largest volumes approach the size of the observable universe. All simulations were performed using the CubeP$^3$M N-body code \citep{HarnoisDeraps:2012he}. 
The CubeP$^3$M code calculates short-range direct particle-particle forces combined with a long-range PM force calculation making it a P$^3$M (particle-particle-particle-mesh) code. It is massively parallel and runs efficiently on either distributed- or shared-memory machines. This is achieved via a cubical, equal-volume domain decomposition combined with a hybrid OpenMP and MPI approach. CubeP$^3$M scales well up to thousands of processors and to date has been run on up to 21,976 computing cores \citep{2008arXiv0806.2887I,2010arXiv1005.2502I,HarnoisDeraps:2012he}. We note that such large simulations 
yield very large amounts of data, with just the particle data amounting to between 700 GB ($3072^3$) and 4.7 TB ($6000^3$) per time-slice, providing significant challenges in the data handling and analysis, as we discuss below.

\begin{table*}
\caption{N-body simulation parameters. Background cosmology 
is based on the WMAP 5-year results. Minimum halo mass and the redshift of formation of the first halo are based on a 20 particle minimum. Resolution here refers specifically to the softening length, above which the interparticle force is the exact Newtonian force (see \protect\cite{HarnoisDeraps:2012he} for details). See text for details regarding the quantities $\mathcal{P}(k)_{max}$ and $\delta_{in}^{rms}/\Delta_p$.
}
\label{summary_N-body_table}
\begin{center}
\begin{tabular}{@{}lllllllllllllllllll}
\hline
Box Size & $\mathrm{N}_{part}$ & Mesh & Resolution & $\mathrm{m}_{particle}$ & $M_{halo,min}$ & $z_{\mathrm{in}}$ & $\mathcal{P}(k)_{max}$ & $\delta_{in}^{rms}/\Delta_p$ & $z_{\mathrm{first halo}}$
\\[2mm]
$ h^{-1}$Mpc & & & $h^{-1}$kpc & $\,h^{-1}\mathrm{M}_\odot$ & $\,h^{-1} \mathrm{M}_\odot$ &  &  &  & 
\\[2mm]\hline
11.4 & $3072^3$ & $6144^3$ & 0.18 & $3.63\times10^3$ & $7.63\times10^4$ & 300 & $2.0\times10^{-5}$ & 0.037 & 41
\\[2mm]
20 & $5488^3$ & $10976^3$ & 0.18  & $3.63\times10^3$ & $7.63\times10^4$ & 300 & $2.0\times10^{-5}$ & 0.045 & 44
\\[2mm]
114 & $3072^3$ & $6144^3$ & 1.86 & $3.83\times10^6$ & $7.63\times10^7$ & 300 & $1.2\times10^{-5}$ & 0.069 & 30
\\[2mm]
425 & $5488^3$ & $10976^3$ & 3.87 & $3.69\times10^7$ & $7.35\times10^8$ & 300 & $9.5\times10^{-6}$ & 0.068 & 25
\\[2mm]
1000 & $3456^3$ & $6912^3$ & 14.47 & $1.96\times10^9$ & $3.85\times10^{10}$ & 150 & $4.6\times10^{-4}$ & 0.057 & 17
\\[2mm]
3200 & $4000^3$ & $8000^3$ & 40.00 & $4.06\times10^{10}$ & $8.12\times10^{11}$ & 120 & $4.5\times10^{-4}$ & 0.052 & 11
\\[2mm]
6000 & $6000^3$ & $12000^3$ & 50.00 & $7.49\times10^{10}$ & $1.50\times10^{12}$ & 100 & $2.8\times10^{-5}$ & 0.051 & 11
\\[2mm]
\hline
\end{tabular}
\end{center}
\end{table*}

\subsection{Cosmology}

We base our simulations on the 5-year WMAP results \citep{Dunkley:2008ie, Komatsu:2008hk}. With the exception of two of our runs (the 1 $h^{-1}$Gpc and 3.2 $h^{-1}$Gpc boxes) the cosmology used for the simulations was the `Union' combination from \cite{Komatsu:2008hk}, based on results from WMAP, baryonic acoustic oscillations and high-redshift supernovae; i.e. $\Omega_{m}=0.27$, $\Omega_{\Lambda}=0.73$, $h=0.7$, $\Omega_{b}=0.044$, $\sigma_8=0.8$, $n_s=0.96$. The 1 $h^{-1}$Gpc and 3.2 $h^{-1}$Gpc boxes were based on the slightly different `Alternative' combination from the same paper; i.e. $\Omega_{m}=0.279$; $\Omega_{\Lambda}=0.721$, $h=0.701$, $\Omega_{b}=0.046$, $\sigma_8=0.817$, $n_s=0.96$. The power spectrum and transfer function used for setting initial conditions was generated using CAMB \citep{Lewis:1999bs}.

\subsection{Initial Redshift and Initial Conditions}

The CubeP$^3$M code uses first-order Lagrangian perturbation theory (1LPT), i.e. the Zel'dovich approximation \citep{1970A&A.....5...84Z}, to place particles in their initial positions. The initial redshift when this step takes place is given for each simulation in table 1. 

There is debate regarding the suitability of the Zel'dovich approximation in setting initial conditions \citep{Crocce:2006ve, Crocce:2009mg, Lukic:2007fc, Tinker:2008ff, Knebe:2009id, Reed:2012ih}. Care must be taken when setting a choice for the initial redshift so as to ensure that artefacts (transients) are minimised. A detailed study by \cite{Lukic:2007fc} emphasises the need for a suitable choice for the initial redshift as applied to a particle-mesh (PM) code \citep[MC$^2$ -- see ][]{Heitmann:2004gz, 2008CS&D....1a5003H} using 1LPT. \cite{Reed:2012ih} recommend the use of second-order Lagrangian perturbation theory (2LPT) in setting initial conditions based on their study using two tree-based codes: Gadget-2 \citep{2005MNRAS.364.1105S} and PKDGRAV \citep{2001PhDT........21S}. 

\cite{Lukic:2007fc} proposed two criteria to guide the choice of initial redshift for simulations employing 1LPT: (1) whether the amplitude of the initial power spectrum modes in the box are in the linear regime, and (2) whether the initial particle displacement is small enough so that the particle grid distortion is relatively small. The first criterion sets a minimum for the initial redshift based on the requirement that the dimensionless power spectrum, $\mathcal{P}(k)=k^3P(k)/2\pi^2$, be less than some arbitrary, small value at the initial redshift (\cite{Lukic:2007fc} used $\mathcal{P}(k)_{max} \le 0.01$). As smaller boxes for a given particle number probe higher values of $k$, earlier initial redshifts are required for smaller boxes. Similarly, for a fixed box size, simulations with greater numbers of particles also require earlier initial redshifts. The second criterion places a more stringent bound on the initial redshift. The example discussed in \cite{Lukic:2007fc} was that the particles were displaced, on average, by no more than $\delta_{in}^{rms}=0.3\Delta_p$, where $\Delta_p=L_{box}/n_p$ is the interparticle spacing and $\delta_{in}^{rms}$ is the root-mean-squared displacement of the particles. In Table 1 the values of $\mathcal{P}(k)_{max}$ and $\delta_{in}^{rms}/\Delta_p$ for our simulations are given. We see that for all of our simulations both of these criteria are satisfied.

\section{Halo finding}
\label{halofining:sect}

\subsection{Motivation}

In the hierarchical picture of structure formation peaks in the linear density field of the early universe grow into highly non-linear structures via gravitational attraction. A popular formalism for describing this process is the identification of the non-linear, high density peaks as dark matter haloes. These haloes, appropriately defined, are concomitant with virialised regions in the universe where galaxies reside. The hierarchical growth of structure in this picture can then be viewed as a series of halo mergers, commencing with very small haloes at early times, and continuing all the way to the present, with haloes growing larger via both mergers and the accretion of smaller haloes. In cosmological simulations it is usual to attempt to track the progress of the growth of structure by identifying haloes within the simulation volume across a range of redshifts.

\subsection{Halo Finding Approaches}

There are two main approaches to finding haloes in cosmological simulations. The first is the spherical overdensity (SO) method. In this method haloes are assumed to be spherical. The extent of a halo is governed by a free parameter, the overdensity criterion, $\Delta$, which is a cut-off in density with respect to some background density (typically either the matter background density or the critical density of the universe). In the SO algorithm spheres are grown from a central location until the enclosed overdensity of the sphere is equal to the overdensity criterion. The main differences between halofinding codes that use this approach lie in how the centres of the candidate haloes are identified.

The second approach is the Friends-of-Friends (FOF) algorithm. This algorithm is also based on one parameter, $b$, the linking length parameter. The algorithm finds haloes that contain particles that are within $b\Delta_p$ of at least one other particle in the halo. While the SO algorithm produces a halo that is by definition spherical, the FOF algorithm creates haloes that are arbitrarily shaped. For a recent comparison project on halofinding codes see \cite{Knebe:2011rx}.

We employ three halo-finding codes in our analysis: CubeP$^3$M's own on-the-fly SO halofinder (hereafter `CPMSO') \citep{HarnoisDeraps:2012he}, the Amiga Halo Finder (hereafter `AHF') \citep{Gill:2004km,Knollmann:2009pb}, and the FOF halofinder from the Gadget-3 N-Body cosmological code (an update to the publicly available Gadget-2 code \citep{2005MNRAS.364.1105S}).

The CPMSO halofinder utilises a fine mesh from the CubeP$^3$M code (a mesh with spacing of $\Delta_p/2$) to identify local peaks in the density field. The code first builds the fine-mesh density using either Cloud-In-Cell (CIC) or Nearest-Grid-Point (NGP) interpolation. It then proceeds to search for and record all local density maxima above a certain threshold (typically set to 100 above the mean density) within the physical volume. It then uses quadratic interpolation on the density field to determine more precisely the location of the maximum within the densest cell. The halo centre determined this way agrees closely with the centre-of-mass of the halo particles. Each of the halo candidates is inspected independently, starting with the highest peak. The grid mass is accumulated in spherical shells of fine grid cells surrounding the maximum until the mean overdensity within the halo drops below $\Delta$. While the mass is accumulated it is removed from the mesh, so that no mass element is double-counted. This method is thus inappropriate for finding sub-haloes as within this framework they are naturally incorporated in their host haloes. Because the haloes are found on a grid of finite-sized cells and spherical shells constructed from them, it is possible, especially for the low-mass haloes, to overshoot the target overdensity. When this occurs we use an analytical halo density profile to correct the halo mass and radius to the values corresponding to the target overdensity. This analytical density profile is given by the Truncated Isothermal Sphere (TIS) profile \citep{1999MNRAS.307..203S,2001MNRAS.325..468I} for overdensities below $\sim130$, and $1/r^2$ for lower overdensities. The TIS density profile has a similar outer slope (the relevant one here) to the Navarro, Frenk and White (NFW) profile \citep{1997ApJ...490..493N}, but extends to lower overdensities and matches well the virialization shock position given by the Bertschinger self-similar collapse solution \citep{1985ApJS...58...39B}. For further details on the CPMSO method see  \cite{HarnoisDeraps:2012he}.

Details for AHF can be found in \cite{Knollmann:2009pb}. The algorithm identifies density peaks using a nested set of grids that are finer-grained in regions of higher density. Haloes are then identified by collecting particles together that are contained within isodensity contours on each level of the grid hierarchy. This allows AHF to identify sub-haloes within host haloes as the algorithm progresses from high resolution grids to lower resolution ones. AHF then removes particles that are unbound and recalculates halo properties based on the remaining bound particles.

The specifics of the FOF halofinder packaged in with the Gadget-3 code currently have not been detailed in any publication but the algorithm itself is outlined in \cite{1985ApJ...292..371D}. The main difference in the algorithm that exists in the Gadget-3 version is that the code is parallelised for distributed-memory machines. Specifically, haloes are found in local subvolumes of the simulation assigned to individual MPI tasks (created using the Gadget-3 domain decomposition which utilises a space-filling Peano-Hilbert curve -- for details see the Gadget-2 paper \cite{2005MNRAS.364.1105S}) and then haloes that extend spatially beyond the edges of the subvolumes are linked together in a final MPI communication step. We have altered the Gadget-3 code to read CubeP$^3$M's particle output format and significantly reduced its memory footprint by stripping away extraneous data structures.

Finally, we note that due to limitations in the scaling of the codes with processor numbers, large memory footprint, and incompatible data structures, in order to apply the AHF and FOF algorithms to our data it was necessary to split the simulation time-slices into a number of subvolumes and run the halofinding algorithms on each subvolume independently. Each subvolume included a buffer zone which overlapped with the neighbouring ones, for correct handling of haloes straddling two or more sub-regions. We then stitched the subvolumes back together to create the final AHF and FOF halo catalogues, removing any duplicated structures in the overlapping buffers. Although somewhat more expensive than applying each halo finder directly on the full data, this approach allows the handling of much larger amounts of data than otherwise possible and provides additional flexibility in terms of computational resources needed for post-processing.

\subsection{Halo Definitions}

In this work we choose the overdensity criteria for identification of spherical overdensity haloes to be $\Delta_{178}$, i.e. an overdensity of 178 times the background matter density. This is a common choice motivated from the top-hat model of non-linear collapse in an Einstein de-Sitter (EdS) universe \citep{1972ApJ...176....1G}. The overdensity criterion is usually taken to be an overdensity with respect to either the background matter density or the background critical density. For an EdS universe, i.e. a universe containing only collisionless matter, the two are the same. For a $\mathrm{\Lambda CDM}$ universe, at late times, the two deviate from each other due to the increasing dark energy component that contributes to the critical density of the universe but not to the matter density. Care must be taken to compare like-for-like overdensities, especially during the $\Lambda$-dominated epoch (note that all references in this work to overdensities refer to those with respect to the background matter density). Any given non-linear overdensity criterion can be mapped onto a threshold value in the linear regime. For $\Delta_{178}$ this corresponds to a linear overdensity of $\delta_c=1.686$ in an EdS universe. In a $\mathrm{\Lambda CDM}$ universe the value of $\delta_c$ evolves to 1.674 at $z = 0$ due to the influence of dark energy. In the case of the FOF halofinder we follow various previous authors \citep{2001MNRAS.321..372J,Reed:2003sq, Reed:2006rw,Crocce:2009mg,Courtin:2010gx,Angulo:2012ep} and use a linking length of 0.2 for our analysis.

\subsection{Halo Mass Redshift Dependence}
\label{hm_redshift_dep}

There are a number of factors that influence the growth of halo mass. They can be divided into `physical' mass growth, i.e. the growth of the haloes via mergers and accretion, and, following the nomenclature of \cite{Diemer:2012mw}, `pseudo' mass growth. For a recent study on the relative effects of physical and pseudo mass growth on the SO mass function between $z=0-1$ see \citet{Diemer:2012mw} and references therein. An important point from this study is that a significant fraction of SO halo mass growth between $z=0-1$ is due to pseudo-evolution of halo mass. This is growth of halo mass, defined in the SO sense, due to the evolution of the background density of the universe. As the background matter density of the universe decreases with the universe's expansion, the radius of a given SO halo will grow as it requires a lesser enclosed physical density to meet the overdensity criterion of $\Delta$ times the background density. In the FOF case there also exists a pseudo mass growth: as the linking length is defined as a constant length in co-moving coordinates its physical length increases as $z$ decreases. The implication of this is that for a halo with a static profile the extent of an FOF halo will increase with time. \cite{2010MNRAS.401.2245F} split FOF halo growth into an accretion component and a diffuse component, the latter containing an element of pseudo growth. Currently there is no equivalent study to the \cite{Diemer:2012mw} work on pseudo mass growth in SO haloes for FOF haloes.

In addition to the pseudo evolution of halo mass there is the evolution of $\delta_c$ (and $\Delta$) due to dark energy dominance at low redshifts. This has well-studied implications for the SO halo mass definition \citep{1991MNRAS.251..128L, 1993MNRAS.262..627L,1996MNRAS.282..263E}. Because of this $\delta_c(z)$ is often used as a parameter in mass functions \citep{1974ApJ...187..425P, 2002MNRAS.329...61S, Reed:2003sq, Reed:2006rw, Courtin:2010gx}. In this work we provide an SO mass function fit for haloes in an $\mathrm{\Lambda CDM}$ universe that is based on $\Delta_{178}$ and do not include $\delta_c(z)$ as a variable in the fit. This is motivated by the fact that we parameterise redshift evolution occurring due to dark energy, pseudo mass growth, and systematic effects such as halo mergers (which occur at different rates in different epochs) and non-sphericity of haloes. These latter effects are not captured by the evolution of $\delta_c(z)$. We also note that a baryonic component will have an effect on the mass function (estimated by \cite{2012MNRAS.423.2279C} to increase the mass function by $\lesssim 3\%$ for a $\Delta_{200}$ halo definition) not captured in this work. 

An issue that pseudo mass evolution raises is that any redshift evolution of the mass function (whether expressed in terms of mass or $\mathrm{ln}\sigma^{-1}$, see \S~\ref{hmf:sect}) is due to an interplay between halo definition and the physical increase of halo mass with time. In order to link halo definitions to observations and compare our results to previous studies on the halo mass function we adopt the common halo definitions above and attempt to capture redshift evolution based on them.

\section{The Halo Mass Function}
\label{hmf:sect}

\begin{figure}
\includegraphics[width=3.3in]{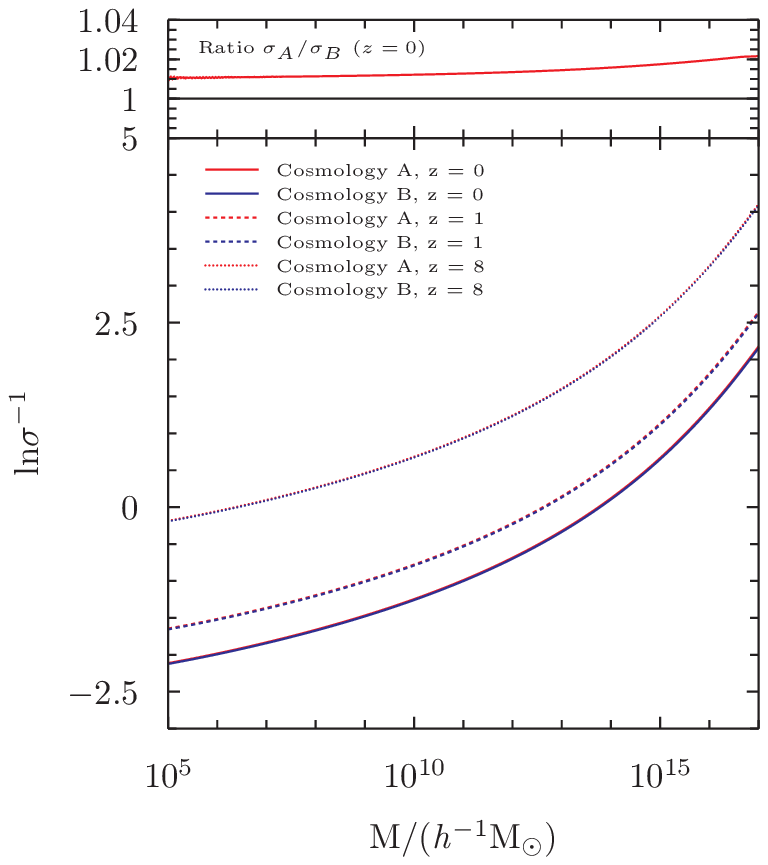}

\captionof{figure}{The relationship between the variance of the linear density field, $\sigma$, and halo mass for two $\Lambda \mathrm{CDM}$ cosmologies. `Cosmology A' refers to the `Union' model of \protect\cite{Komatsu:2008hk} with $\Omega_{m}=0.27$, $\Omega_{\Lambda}=0.73$, $h=0.7$, $\Omega_{b}=0.044$, $\sigma_8=0.8$, $n_s=0.96$. `Cosmology B' refers to the `Alternative' model of \protect\cite{Komatsu:2008hk} with $\Omega_{m}=0.279$; $\Omega_{\Lambda}=0.721$, $h=0.701$, $\Omega_{b}=0.046$, $\sigma_8=0.817$, $n_s=0.96$.}
\label{fig:mass_sigma}
\end{figure}

The mass function can be expressed in a number of ways. Here we use the multiplicity function, $f(\sigma,z)$, which represents the fraction of mass that has collapsed to form haloes per unit interval in ln$\sigma^{-1}$. This definition uses $\sigma^{-1}$ as a proxy for mass, where $\sigma^{2}$ is the variance of the linear density field, given by:

\be
\sigma^2(M,z)=\frac{D^2(z)}{2\pi^{2}}\int_0^\infty k^2P(k)W^2(k;M)\mathrm{d}k
\ee
where $P(k)$ is the power spectrum of the linear density field, $W(k;M)$ is the Fourier-space representation of a real-space top-hat filter containing mass $M$ -- assuming that the top-hat sphere encloses a region that contains a mean density identical to that of the universe -- and $D(z)$ is the growth factor, normalised to the unity at $z = 0$ \citep{1993ppc..book.....P}. The radius in real-space of the filter can be set using an overdensity criterion, $\Delta$, as per the SO algorithm. The relationship between $\sigma$ and halo mass for our two WMAP-5-based cosmologies is shown in Figure~\ref{fig:mass_sigma}.

We define the halo multiplicity function as \citep{2001MNRAS.321..372J}:

\be
f(\sigma,z) \equiv \frac{M}{\rho_m(z)}\frac{dN(M,z)}{d\mathrm{ln}\sigma^{-1}} 
\label{mult_funct}
\ee
where $N(M,z)$ is the mass function proper, that is to say the number count of haloes with a mass less than $M$ per unit volume, and $\rho_m$ is the mean matter density of the universe.

\subsection{Mass Binning}

The halofinders calculate values for the masses of all resolved haloes. It is then necessary, in order to construct $f(\sigma,z)$, to count the number of haloes in mass bins before converting the masses into values of $\mathrm{ln}\sigma^{-1}$. Rather than equation 2, for simulated halo catalogues we have:

\be
f(\sigma,z) \equiv \frac{M^2}{\rho_m(z)}\frac{\Delta N_{sim}}{\Delta M}\frac{d\mathrm{ln}M}{d\mathrm{ln}\sigma^{-1}} 
\ee
where care must be taken in choosing the width of the bins in mass as this can potentially result in a source of systematic error. For a detailed evaluation of this error see \cite{Lukic:2007fc}, who note that the error is not significant so long as the bin widths, $\Delta \mathrm{log}M$, do not exceed 0.5. We use bin widths for our analysis that remain constant in $\mathrm{log}{M}$ with $\Delta \mathrm{log}M$=0.16. and assign the mass of a given bin to the average of all the haloes in the bin rather than the bin centre.

\subsection{Error Treatment}
\label{sect:error}

Throughout this paper we use 1-$\sigma$ Poisson error bars as defined in \cite{Heinrich:2003q}:

\be
\sigma_\pm=\sqrt{N_{haloes}+\frac{1}{4}} \pm \frac{1}{2}
\ee
The main advantages of using error bars in this form are: (1) for small numbers of haloes the asymmetry of the bars reflects the asymmetry of the Poisson distribution; (2) for a halo count of 1 the lower edge of the error bar does not reach zero. Note that for large numbers of haloes the errors tend to the expected $\sqrt{N}$ form.

\subsection{Data Treatment}

In order to combine data across our simulations certain systematic effects need to be accounted for. These include: accounting for the finite volumes of our simulations; adjusting the FOF haloes for a systematic overestimation in mass for haloes sampled with low particle numbers; the question of whether to remove or include sub-haloes from our AHF data; and the choice of a lower limit to the number of particles a halo contains. The mass functions presented here are (1) corrected for finite volumes for simulations with a box size of less than 425 $h^{-1}$Mpc, (2) adjusted, in the case of our FOF haloes, to account for mass overestimation using the empirical correction of \cite{Warren:2005ey}, (3) constructed based only on host haloes (in the case of data from the AHF halofinder the host haloes have masses that include any sub-haloes they contain) and, (4) have a minimum of 1000 particles in each halo. We now discuss each of these points in detail.

\subsubsection{Finite Volume Correction}

The volume of space modelled by a cosmological simulation is always finite. However when we speak of the mass function we typically refer to a `global' mass function, i.e. one that would correspond to an infinitely large volume. There is therefore a disconnect that needs to be bridged between simulations and an ideal, global, mass function \citep{Sirko:2005uz,Bagla:2006ym,Power:2005ie}. There exist a number of ways to address this issue. In this work we adopt an approach recently employed by \cite{Lukic:2007fc} and \cite{Bhattacharya:2010wy}. We proceed by assuming that mass function universality holds strictly in the sense that the functional form is the same for both global and local volumes. This is similar to assuming that the mass function is universal across cosmologies that differ in their mass-$\sigma$ relations, as the effect of a finite box size is to set to zero the amplitude of any density fluctuations on a scale greater than the box size. There is evidence that the mass function is not universal (see \S~\ref{summary:sect}); that its functional form has a weak dependence on both redshift and cosmology. Therefore any finite volume correction we make based on the assumption of universality is an approximation. We adopt the approach here despite this, as it is relatively straight-forward to apply and it brings our data into better agreement across difference box sizes. We also briefly discuss below other possible methods that could be applied.

\begin{figure}
\includegraphics[width=3.3in]{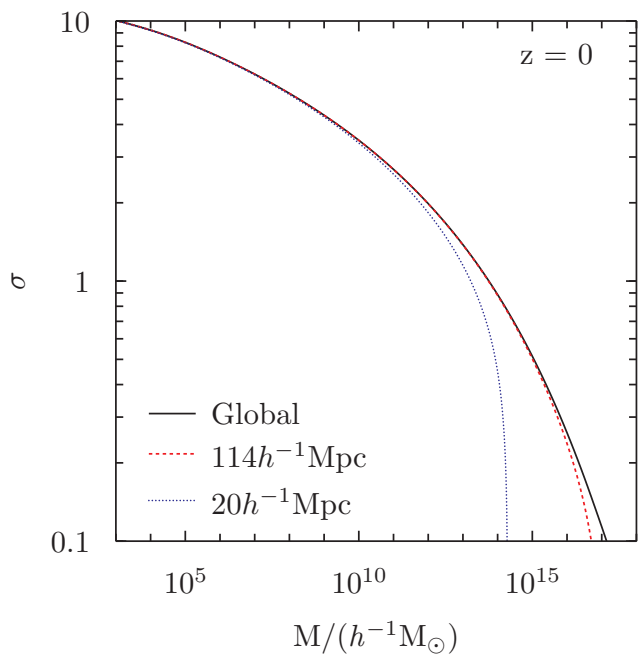}
\captionof{figure}{The finite volume correction for small boxes. The global $M-\sigma$ relation is adjusted to give a relation based on the volume of the simulation box in question via equation~\ref{sigma_correction}. The (approximate) assumption of mass function universality is then employed to adjust the masses and numbers of dark matter haloes via equations~\ref{dn_correction} and \ref{sigma_correction}.}
\label{fig:box_cor_sigma}
\end{figure}

The approach is as follows. Assuming equation~\ref{mult_funct} refers to the mass function in an infinite simulation volume we can re-write it as:
\be
\frac{dN}{dM}=\frac{\rho_m}{M^2}f(\sigma)\frac{d\mathrm{ln}\sigma^{-1}}{d\mathrm{ln}M}
\label{mf1}
\ee
and now for a finite simulation volume we can write:
\be
\frac{dN'}{dM'}=\frac{\rho_m}{M'^2}f(\sigma)\frac{d\mathrm{ln}\sigma'^{-1}}{d\mathrm{ln}M'}
\label{mf2}
\ee
where now $\sigma'(M')$ is determined by the discrete power spectrum of the simulation in question. Also $M'$ is a function of $M$ defined such that $\sigma(M) \equiv \sigma'(M'(M))$. The assumption that the mass function is universal allows us to say that the multiplicity functions, $f$, in equations~\ref{mf1} and \ref{mf2} are identical. This then leads to the relation:

\be
dN=dN'\frac{dM'(M)}{dM}
\label{dn_correction}
\ee
We now require some method of connecting $\sigma'(M')$ and $\sigma(M)$. The extended Press-Schechter formalism \citep{1991ApJ...379..440B} approximately connects $\sigma'(M')$ and $\sigma(M)$ via:

\be
\sigma'(M')^2=\sigma(M)^2 - \sigma^2_{R(box)}
\label{sigma_correction}
\ee
where $\sigma^2_{R(box)}$ is the variance of fluctuations in spheres that contain the simulation volume. To align this to the simulations we make another approximation and equate the spherical volume to the cubical simulation volume. The resulting mass-$\sigma$ relations at $z=0$ for the 114 $h^{-1}$Mpc and 20 $h^{-1}$Mpc boxes are shown in Figure~\ref{fig:box_cor_sigma}.

The steps in the volume correction are then as follows: (1) Calculate $\sigma^2_{R(box)}$ for the simulation box in question. (2) For each bin in $\sigma$ calculate an adjusted bin $\sigma$ value, $\sigma_{cor}$, using $\sigma_{cor}^2=\sigma^2 - \sigma^2_{R(box)}$. (3) Adjust each halo count using equation~\ref{dn_correction}, where $M$ is the mass that corresponds to $\sigma$ and $M'$ the mass that corresponds to $\sigma_{cor}$.

For comparative purposes we now discuss other solutions that have been employed to solve the finite volume issue. \cite{Reed:2006rw} performed a number of different N-body simulations and calculated the range of $\sigma$ for each one using the input power spectra of the simulations and the relationship between $\sigma$ and mass for a finite box in the discrete case:
\be
\sigma^2(M,z)=D^2(z)\sum_\mathbf{k} |\delta_\mathbf{k}|^2W^2(k;M)
\ee
where $|\delta_\mathbf{k}|$ is the linear amplitude of the Fourier modes in the simulation at $z=0$. This approach has the advantage of both correcting for finite volumes and also compensating for cosmic variance. One drawback to it though is that each realisation has a different mass-$\sigma$ relationship. We did not adopt this approach as it is a method more suited to multiple simulation runs, where the mass-$\sigma$ relations can be averaged over to produce a mass function fit. We did not have the luxury of repeating our runs due to the large sizes of the simulations. \cite{Bagla:2009mg} choose to make no specific correction (although they remove any data points that are affected by by more than a threshold level of error -- $10\%$ in the number counts -- from their analysis in a similar manner to \cite{2008ApJ...688..709T}). This decision was motivated by the observation that the mass function is an unknown function that is deduced from simulation data so to place any a priori constraints upon it -- such as a universal functional form for global and local mass-$\sigma$ relations -- is undesirable. This approach was applicable in the study undertaken by \cite{Bagla:2009mg} as the box sizes used for their simulations were all relatively large ($\ge$ 256 $h^{-1}$Mpc), for our smaller boxes not correcting for finite volumes would lead to undesirable systematic discontinuities across our simulations. \cite{Yoshida:2003wf} and \cite{Bagla:2006ym} replaced equation 1 with:

\be
\sigma_{box}^2(M,z)=\frac{D^2(z)}{2\pi^{2}}\int_{2\pi/L}^\infty k^2P(k)W^2(k;M)\mathrm{d}k
\ee
which takes a cut-off in low k modes at the size of the box. \cite{Lukic:2007fc} note that a correction of this form has a dependence on the accuracy of the mass functions at redshifts greater than $z=5$. They also note that in comparison to their correction method it exhibits offsets and shape changes across different box sizes. Finally, \cite{2004ApJ...609..474B} utilised the extended Press-Schechter formalism to create a volume adjustment by taking the Sheth-Tormen mass function for a global volume and applying a correction to it to create a mass function suitable for a smaller volume. As we are not adopting a mass function in the Sheth-Tormen form this approach was inappropriate here.

\subsubsection{Warren Correction to FOF Haloes}

\cite{Warren:2005ey} proposed a simple correction to the masses of FOF haloes that alleviates a systematic error in halo masses calculated via the FOF algorithm at low particle counts. This correction was devised based on analysis of FOF haloes at $z=0$, but has been checked by \cite{Lukic:2007fc} for FOF haloes at higher redshifts. The FOF algorithm overestimates the masses of haloes when there are low numbers of particles sampling the haloes. \cite{Warren:2005ey} proposed the following correction to particle counts in haloes:
\be
N_{corrected}=N(1-N^{-0.6})
\ee

We have adopted this correction in this work and all FOF data shown includes it. As we have also adopted a cut-off in particle number of 1000 the maximum effect this correction has on the masses of our FOF haloes is $\sim 2\%$. 

There is debate regarding the appropriate correction to use. \cite{Bhattacharya:2010wy} find that a correction of $N_{cor}=N(1-N^{-0.65})$ is more suitable and \cite{Lukic:2008ds} note that the halo concentration parameter for haloes with an NFW profile also affects the correction, a result that is corroborated by \cite{More:2011dc}. The more sophisticated corrections proposed by \cite{Lukic:2008ds} and \cite{More:2011dc} (the former provide a correction for FOF haloes with $b=0.2$ whereas the latter provide a correction valid for different values of $b$) are based on individual halo concentrations. We have not adopted this approach as we do not possess profile information for our individual FOF haloes. The difference between the Warren correction and the \cite{More:2011dc} correction (which can be as much as 15\% for haloes with lower particle counts) becomes slight for haloes with many particles. For our smallest, 1000 particle haloes, the correction to the Warren formula is of the order of $<5\%$ depending on halo concentrations. Not accounting for this therefore introduces a small systematic error which in the very worst case of 1000 particle haloes would alter the FOF masses by $\sim 0.15\%$. In addition \cite{More:2011dc} note that the Warren correction is specifically calculated to correct for the bias in the mass function itself whereas their study is one that corrects for the masses of individual FOF haloes when considered in isolation.

\subsubsection{Sub-Haloes}

As noted in \S~\ref{halofining:sect} the AHF halofinder produces halo catalogues that contain sub-haloes -- bound structures inside host haloes. For the purposes of constructing a mass function we have to make a choice as to how to deal with these haloes. The CPMSO and FOF halofinders both identify host haloes only, therefore for compatibility we make the natural choice of excluding the sub-haloes from the mass functions based on the AHF catalogue. The host haloes detected by AHF have properties, including mass, that are calculated based on all the bound matter contained within the halo, including the sub-haloes themselves. This results in the AHF host haloes being directly comparable to the CPMSO ones.

\begin{figure}
\includegraphics[width=3.3in]{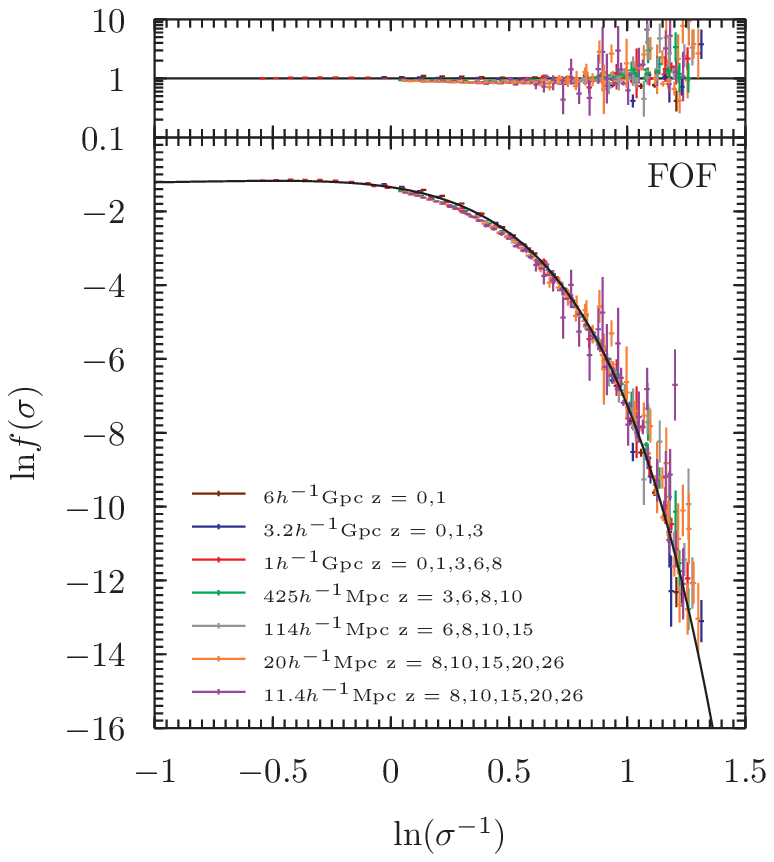}

\captionof{figure}{The FOF mass function across all simulations and redshifts along with our fit (equation~\ref{fof_fit}). The top panel shows the ratio of the data to the fitting formula.}
\label{fig:fof_all}
\end{figure}

\subsubsection{Low-End Particle Cutoff}

We take a minimum particle cut-off of 1000 particles per halo in this study. This is conservative as often halo-finders report haloes with as little as 20 particles, or even fewer. \cite{Lukic:2007fc} recommend a 300 particle minimum for constructing a mass function with a pure PM code. \cite{Warren:2005ey} take an `aggressive' 400 particle minimum in their study and \cite{Tinker:2008ff} a `conservative' 400. It has been observed recently by \cite{Reed:2012ih} that a minimum particle cut-off of at least 1000 is appropriate. Although this latter figure was quoted with tree-based N-body codes in mind -- rather than a P$^3$M code -- we use it to motivate our cut-off of 1000 particles. We note that for studies in other areas it might be  acceptable to use haloes with lower particle counts, e.g. in clustering studies we need to reliably identify and locate a halo, but not necessarily know its mass or other properties precisely. For mass function calculations it is important that haloes are assigned masses that are as close to their correct values as possible, and more particles are required to ensure this is the case.

\subsection{The Universal FOF Function}

We find that the halo mass function based on our FOF halo catalogues follows an universal fit applicable for data from all simulations across all redshifts. The fit is good to within $\sim 10\%$ for most data points (Figure~\ref{fig:fof_all}) and takes following the form:
\be
f(\sigma)=A\left[\left(\frac{\beta}{\sigma}\right)^\alpha+1\right]e^{-\gamma/\sigma^2}
\label{fof_fit}
\ee
where $A=0.282$, $\alpha=2.163$, $\beta=1.406$, $\gamma=1.210$. This fit is valid in the range: $-0.55 \le \mathrm{ln}\sigma^{-1} < 1.31$, which at $z=0$ corresponds to haloes with masses between $1.8\times10^{12}$ and $7.0\times10^{15}h^{-1}\mathrm{M}_{\odot}$. Our largest halo, found in the 3.2 $h^{-1}$Gpc box at $z=0$ has a mass of $6.4\times10^{15}h^{-1}\mathrm{M}_{\odot}$ and $\mathrm{ln}\sigma^{-1}=1.17$, whereas our highest $\mathrm{ln}\sigma^{-1}$ value comes from a $z=3$ halo in the same simulation, with a mass of $1.4\times10^{14}h^{-1}\mathrm{M}_{\odot}$. It should be noted that this extreme halo is most likely the result of the FOF algorithm linking together two large haloes via a bridge of particles -- a systematic effect known as overlinking, discussed in \S~\ref{summary:sect} below. At higher $\mathrm{ln}\sigma^{-1}$ values, i.e. for high mass/high redshift, rare haloes, the scatter about the fit increases dramatically due to shot noise.

Whilst it is a remarkable result that over such a large range of $z$ and $\sigma$ we observe a halo mass function that approximately conforms to a universal shape we urge caution as the mass function is not completely universal and exhibits a modest redshift evolution (see \S~\ref{summary:sect} for details).

\begin{figure}
\includegraphics[width=3.3in]{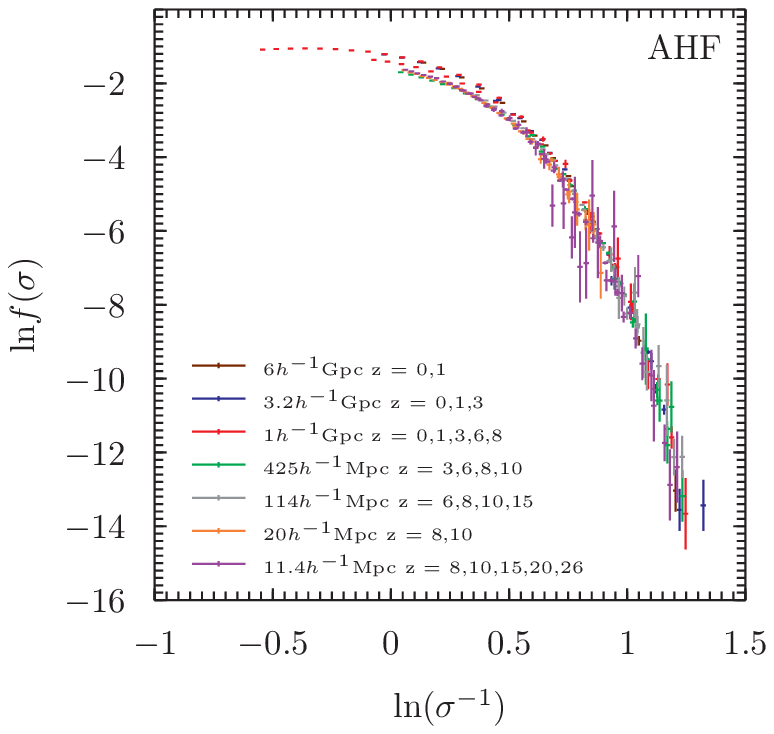}
\captionof{figure}{The AHF mass function across all simulations and redshifts.}
\label{fig:ahf_all}
\end{figure}

\subsection{Mass Function for Spherical Overdensity Haloes}

\subsubsection{Redshift Evolution}
\label{MF:so_redshift_ev}

\begin{figure}
\includegraphics[width=3.3in]{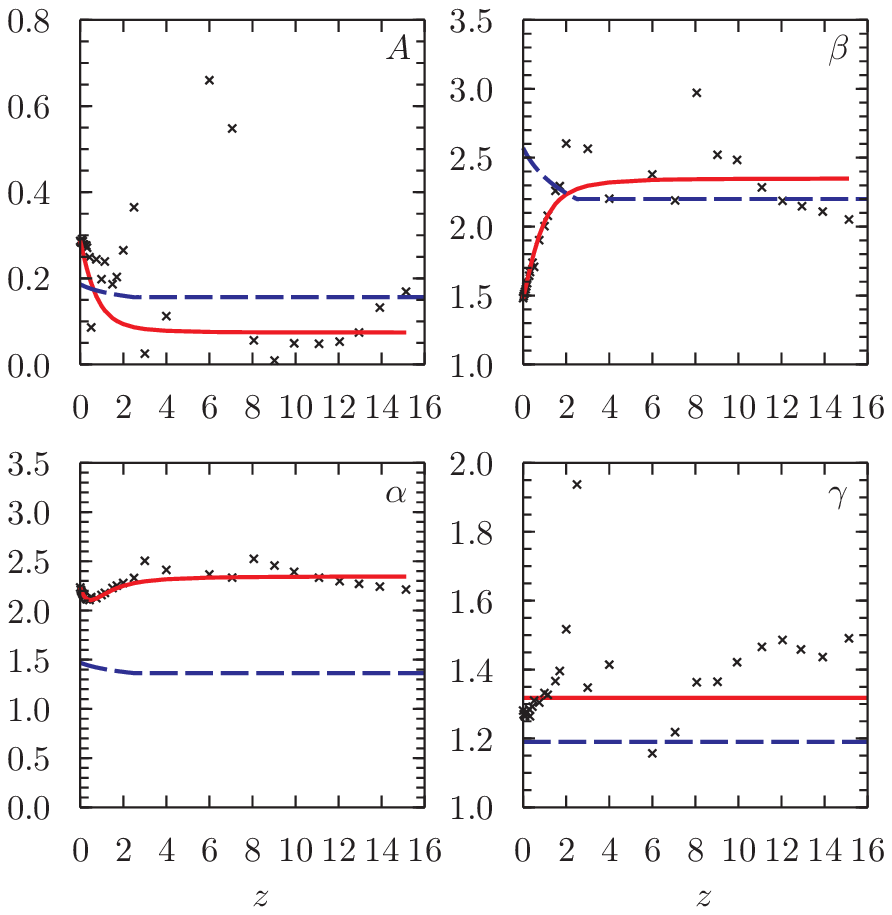}

\captionof{figure}{Evolution of the parameters in equation~\ref{fof_fit} for CPMSO haloes. The \protect\cite{Tinker:2008ff} values are shown in blue for comparison, the model used in this work is shown in red. The parameters have been fitted in the order: $\gamma\to A \to \alpha \to \beta$.}
\label{fig:cpm_param_z_dep_all}
\end{figure}

\begin{figure*}
\includegraphics[width=7in]{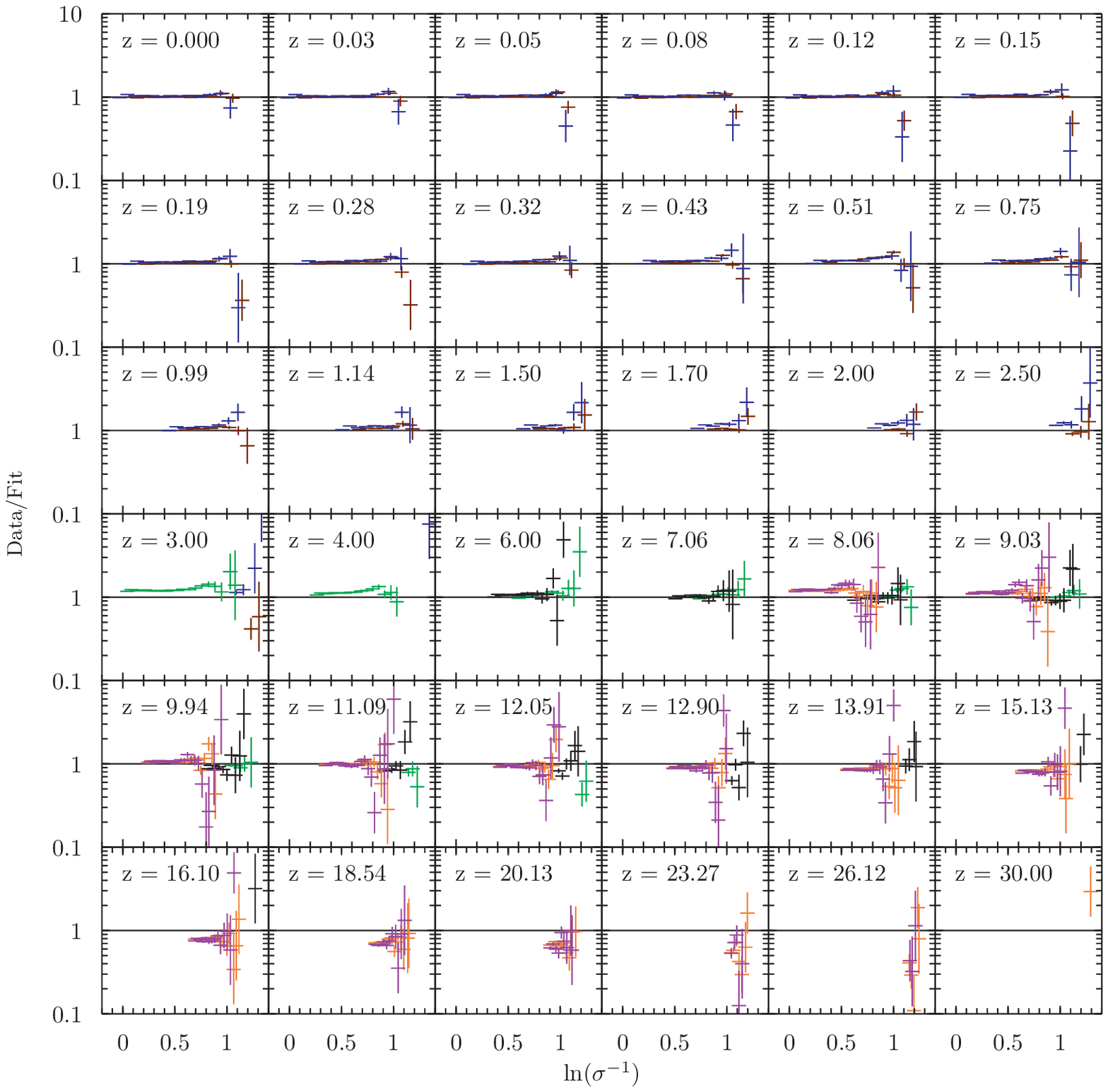}

\captionof{figure}{Ratios between the CPMSO redshift-dependent mass function and the data from the CPMSO halofinder across 36 redshifts from $z=0-30$. The colour scheme is as per Figure~\ref{fig:fof_all}.}
\label{fig:multi_res_z_dep_cpm_all}
\end{figure*}

We show in Figure~\ref{fig:ahf_all} the mass function from our AHF haloes. In contrast to the FOF mass function it is clear that a universal fit is not appropriate for spherical overdensity haloes (a fact also born out in the results from the CPMSO halofinder. For a comparison between the CPMSO and AHF halo mass functions see Figure~\ref{fig:cpm_vs_ahf} and discussion below). We again adopt the fitting function in equation~\ref{fof_fit}, but with a parameterisation that includes a redshift-dependence, as discussed below. Since running AHF on multiple checkpoints from our largest simulations is computationally expensive we utilise the data from the CPMSO on-the-fly halofinder in order to investigate the redshift-dependence of the parameters for the spherical overdensity case. In \S~\ref{halo_finder_comparisons:sect} below we discuss the differences between the three halo finding methods and re-scale the fit to match the AHF results. We perform a least-squares fit on the CPMSO data for 36 output redshifts, initially allowing all of the parameters in equation~\ref{fof_fit} to vary. \citet{Tinker:2008ff} previously found, based on lower-redshift data ($0<z<2$), that the parameter controlling the exponential cut-off scale, $\gamma$, is approximately constant across their range of redshifts. We similarly find that $\gamma$ is approximately constant, albeit at a slightly different value from \citet{Tinker:2008ff} (see Figure~\ref{fig:cpm_param_z_dep_all}, bottom-right panel). Given this result we proceed by fixing this parameter at its value in the approximation of universality: i.e. $\gamma=1.318$. We then fit the CPMSO data across all redshifts again. We find that using a parameterisation that includes $\Omega_m(z)$, the matter content of the universe at a given redshift, enables us to capture late time behaviour that differs from that at high-redshift. This gives us a fit for $A(z)$ in the following form:
\be
\begin{array}{lcl}
A(z)=\Omega_m(z)\left\{0.990\times(1+z)^{-3.216} + 0.074\right\}
\end{array}
\label{A:equ}
\ee
With $\gamma$ and $A$ modelled we then repeat the procedure for $\alpha$ and then $\beta$. We find for $\alpha(z)$:
\be
\begin{array}{lcl}
\alpha(z)=\Omega_m(z)\left\{5.907\times(1+z)^{-3.599} + 2.344\right\}
\end{array}
\label{alpha:equ}
\ee
and for $\beta(z)$:
\be
\begin{array}{lcl}
\beta(z)=\Omega_m(z)\left\{3.136\times(1+z)^{-3.058} + 2.349\right\}
\end{array}
\label{beta:equ}
\ee
We show how our model compares to the parameter fitting data in Figure~\ref{fig:cpm_param_z_dep_all}.

In Figure~\ref{fig:multi_res_z_dep_cpm_all} we compare our redshift-dependent fit to our CPMSO data across all redshifts from $z=0$ and $z=30$. We find that the fit is excellent at low redshifts and remains quite good, within 20\%, all the way to $z\sim20$. For $z>20$ our fit still gives a reasonable match, but the data has large error bars due to the scarsity of haloes then. Nonetheless, we note that around $z=3-4$ our fit slightly under-predicts the abundances of SO haloes across the lower-mass $\mathrm{ln}\sigma^{-1}$ range we cover. There is also an apparent under-prediction for lower $\mathrm{ln}\sigma^{-1}$ around $z=8$ and there is perhaps an over-prediction at very high redshifts ($z>15$).

\subsubsection{AHF-based Fits}
\label{halo_finder_comparisons:sect}

\begin{figure}
\includegraphics[width=3.3in]{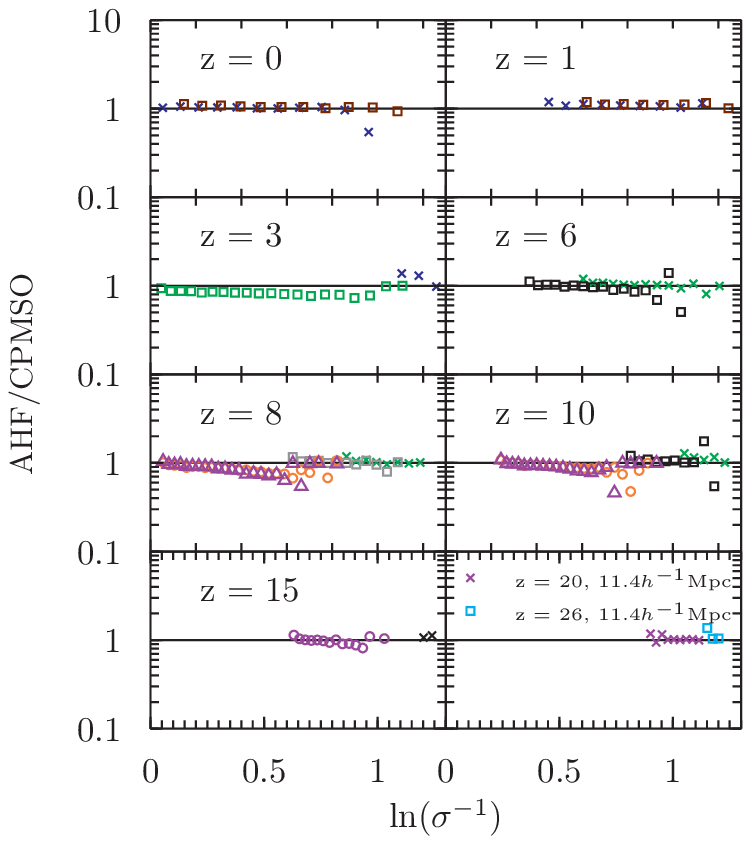}

\captionof{figure}{Comparison of the mass functions from the CPMSO and AHF halofinders across redshifts from $z=0-26$. The colour scheme matches the simulations as per Figure~\ref{fig:fof_all}}
\label{fig:cpm_vs_ahf}
\end{figure}

\begin{figure}
\includegraphics[width=3.3in]{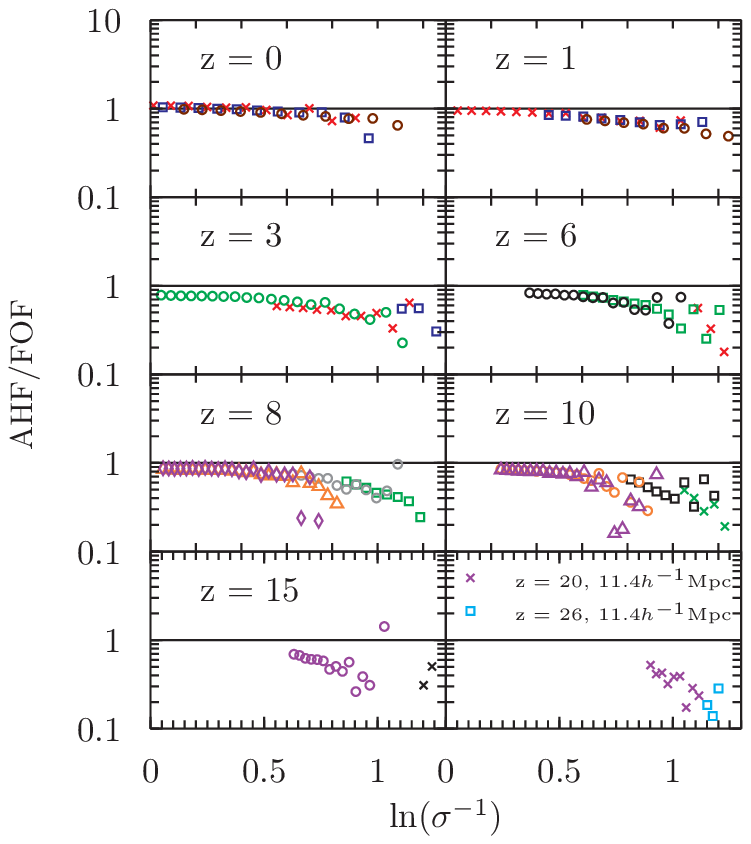}

\captionof{figure}{Comparison of the mass functions from the FOF and AHF halofinders across redshifts from $z=0-26$. The colour scheme matches the simulations as per Figure~\ref{fig:fof_all}}
\label{fig:fofcor_vs_ahf}
\end{figure}

Our CPMSO on-the-fly halo finder is by its nature simplified and potentially more approximate than AHF. In Figure~\ref{fig:cpm_vs_ahf} we show the ratio between the mass functions derived using these two SO-based halofinders, AHF and CPMSO. In  Figure~\ref{fig:fofcor_vs_ahf} we show the ratio between the FOF mass function and the AHF mass function. Both AHF and FOF exhibit consistent behaviour across all simulation volumes and resolutions and for all redshifts. The CPMSO halo finder largely agrees fairly well with AHF, typically within 10-20\%, and much less at low redshifts. There are some systematic differences around $z=8$ in the transition between the 20 $h^{-1}$Mpc and 114 $h^{-1}$Mpc boxes. On the other hand, while both FOF and AHF are consistent across all box sizes, there is a systematic trend for the FOF to yield more rare, massive haloes than AHF for all volumes and redshifts. A similar trend was noted previously by \citet{Reed:2006rw} and \citet{Tinker:2008ff}. Based on these results, we conclude that the AHF (and FOF) data is more consistent across different box sizes than the CPMSO halofinder. 

Given these differences between the two SO halo finders, in addition to the CPMSO fit above, we also provide AHF fits to the data, with three different types: 1) a redshift-dependent one, based around the redshift evolution we have observed in the CPMSO mass function, but re-normalized to the AHF data; 2) a present-day one, based on our results at $z=0$, for direct comparison with previous works; and 3) a more precise high-redshift, ``Epoch of Reionisation (EoR)'' halo mass function, based on all the AHF data past $z=6$.

Given the relationship between the two SO halofinders in Figure~\ref{fig:cpm_vs_ahf} we expect that the redshift parameterisation presented in equations~\ref{A:equ}--\ref{beta:equ} applied to AHF data leads to a mass function that is roughly correct. The fit to AHF at $z=0$ can be improved by introducing a slightly different parameterisation for $A$, as follows:
\be
\begin{array}{lcl}
A(z)=\Omega_m(z)\left\{1.097\times(1+z)^{-3.216} + 0.074\right\}
\end{array}
\label{A_ahf:equ}
\ee

\begin{figure}
\includegraphics[width=3.3in]{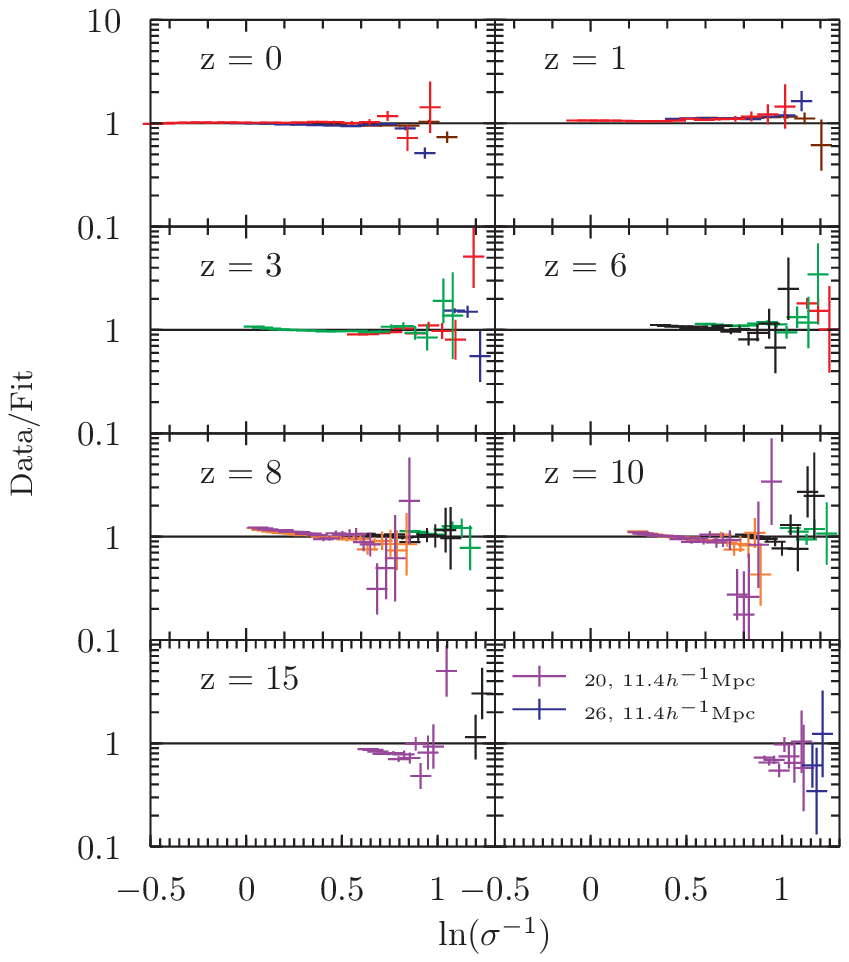}

\captionof{figure}{Ratios between the amplitude-modified CPMSO mass function and AHF data from $z=0$ to $z=26$. The colour scheme is as per Figure~\ref{fig:fof_all}.}
\label{fig:multi_res_z_dep_ill_ahf}
\end{figure}

The combination of equations~\ref{alpha:equ},~\ref{beta:equ},~\ref{A_ahf:equ}, and $\gamma = 1.318$ results in the ratios shown in Figure~\ref{fig:multi_res_z_dep_ill_ahf}. We see that this fit is accurate to $\sim10\%$ for redshifts less than $z=15$. The shape of the function in the low $\mathrm{ln}\sigma^{-1}$ range is slightly different for $z>8$, although the amplitude and the shape are both correct for $\mathrm{ln}\sigma^{-1} > 0.3$ at $z = 8-10$. 

\begin{figure}
\includegraphics[width=3.3in]{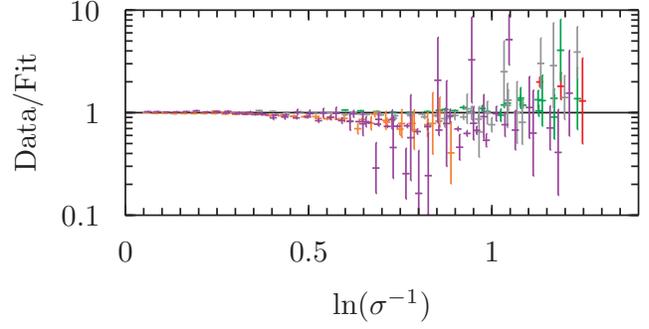}

\captionof{figure}{Ratio between the EoR AHF fit and simulation data from $z=6-26$. The colour scheme is as per Figure~\ref{fig:fof_all}.}
\label{fig:MF_6_plus_ahf}
\end{figure}

We now provide more accurate parameterisations for our AHF data for several different redshift ranges. At $z=0$ we have: $A=0.194$, $\alpha=2.267$, $\beta=1.805$ and $\gamma=1.287$ valid in the range $-0.55 \le \mathrm{ln}\sigma^{-1} < 1.05$. As the redshift evolution of the spherical overdensity mass function is mainly apparent at later times ($z<3$) we can provide an universal fit that is appropriate for high redshift studies, for example for probing the EoR and the Cosmic Dark Ages. An AHF mass function fitted to all our data from $z=6$ upwards has the following parameterisation: $A=0.563$, $\alpha=3.810$, $\beta=0.874$ and $\gamma=1.453$ valid in the range $-0.06 \le \mathrm{ln}\sigma^{-1} < 1.24$ (corresponding at redshift $z = 8$ to a mass range of $3.5\times10^{6}$ -- $6.3\times10^{11}h^{-1}\mathrm{M}_{\odot}$). These values are notably different to those given by our $z = 0$ redshift parameterisation, due to the lack of constraining data for low values of $\mathrm{ln}\sigma^{-1}$. The ratio of the EoR fit versus our data is shown in Figure~\ref{fig:MF_6_plus_ahf}. There is considerable scatter in the high ln$\sigma^{-1}$ end due to data being incorporated from high redshifts when haloes are scarce. For the range $0.05 \le \mathrm{ln}\sigma^{-1} < 0.5$ the fit is accurate to within 10\%.

\subsection{Varying $\Delta$}

\begin{figure}
\includegraphics[width=3.3in]{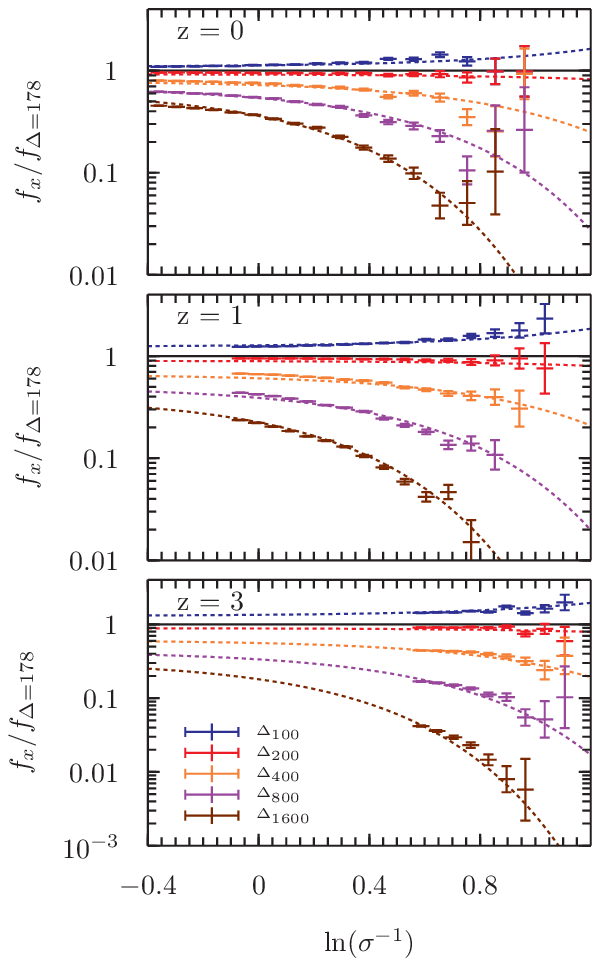}

\captionof{figure}{The effect of the choice of the overdensity criterion, $\Delta$, on the mass function. For z = 0,1,3 (top panel to bottom respectively) the ratio of mass functions with a variety of choices for $\Delta$ are shown relative to the $\Delta_{178}$ function. All data has been calculated from the 1$h^{-1}$Gpc simulation. Fitted curves, described in the text, are shown as dashed lines.}
\label{fig:delta_var_multi}
\end{figure}

For practical purposes it is useful to have a simple method for adjusting the spherical overdensity mass function fits derived above to account for haloes defined with different overdensities. To this end we have run AHF on the 1$h^{-1}$Gpc box using values for $\Delta$ ranging between 100 to 1600, at $z = 0$, 1 and 3. The data has then been fitted with $\Delta$ incorporated into the parameterisation. This allows, to good precision at lower redshifts, a mass function based on a given $\Delta$ to be inferred from our $\Delta_{178}$ fits. The results from this procedure are shown in Figure~\ref{fig:delta_var_multi}.

The parameterisation is as follows. We assume that the $f_{\Delta = 178}$ mass function is suitable for describing mass functions with different choices for $\Delta$ when it is suitably scaled by a function, $\Gamma(\Delta,\sigma,z)$:

\be
\begin{array}{lcl}
f_{\Delta}=\Gamma(\Delta,\sigma,z)f_{\Delta = 178}
\end{array}
\label{Gamma_f:equ}
\ee

\noindent We find that the following form for $\Gamma$ is suitable for describing our data:

\be
\begin{array}{lcl}
\Gamma(\Delta,\sigma,z)=C(\Delta)\left(\frac{\Delta}{178}\right)^{d(z)}\mathrm{\exp}\left[p\left(1-\frac{\Delta}{178}\right)/\sigma^{q}\right]
\end{array}
\label{Gamma:equ}
\ee
 
\noindent Where:

\be
\begin{array}{lcl}
C(\Delta) = \mathrm{ exp}\left[0.023\left(\frac{\Delta}{178}-1\right)\right]\\[2mm]
d(z) = -0.456\Omega_{m}(z)-0.139\\[2mm]
p = 0.072\\[2mm]
q=2.130\\[2mm]
\end{array}
\label{Gamma_param2:equ}
\ee

\noindent The redshift-dependence of the fit is based solely on $\Omega_{m}(z)$, via the $d(z)$ parameter.

\section{Summary and Discussion}
\label{summary:sect}

\subsection{Fitted Functions}

We have provided a number of fitted mass functions based on our data. These are summarised in Table~\ref{summary_fit_table}.

\begin{table}
\caption{Mass function fitting parameters. For details on the redshift evolution of these parameters see sections~\ref{MF:so_redshift_ev} and~\ref{halo_finder_comparisons:sect} 
}
\label{summary_fit_table}
\begin{center}
\begin{tabular}{@{}lllllll}
\hline
 & FOF  & CPMSO & CPMSO & AHF & AHF
\\[0.5mm]
&\textit{Uni.} &&+AHF&&\textit{EoR}
\\[2mm]
& $z = all$ & $z = 0$ & $z = 0$ & $z = 0$ & $z = 6+$
\\[2mm]
\hline
A & 0.282 & 0.287 & 0.316 & 0.194 & 0.563
\\[2mm]
$\alpha$ & 2.163 & 2.234 & 2.234 & 1.805 & 3.810 
\\[2mm]
$\beta$ & 1.406 & 1.478 & 1.478 &  2.267& 0.874
\\[2mm]
$\gamma$ & 1.210 & 1.318 & 1.318 & 1.287 & 1.453
\\[1mm]
\hline
\textit{z dep.} & No & Yes & Yes & No & No
\\[1mm]
\hline
\end{tabular}
\end{center}
\end{table}

\subsection{Comparison to Existing Fits}
\label{sect:comp}

\begin{figure}
\includegraphics[width=3.3in]{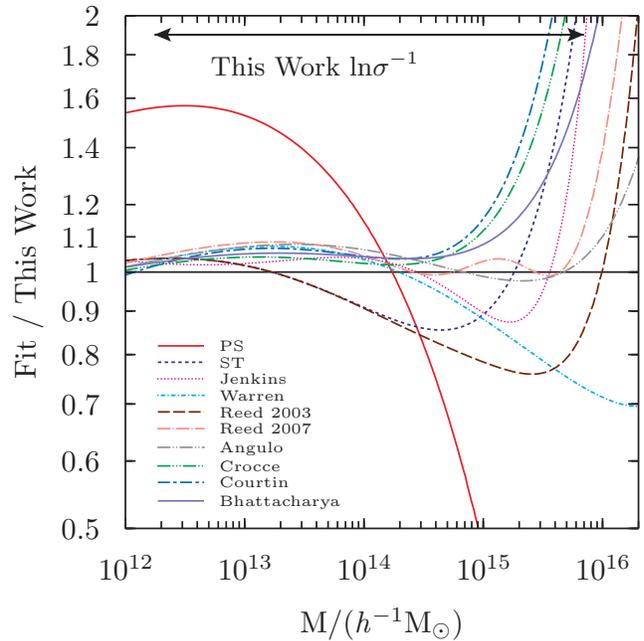}

\captionof{figure}{Ratios between our universal fit for FOF haloes and a number of FOF fits from the literature for $z=0$.}
\label{fig:analytic_comp_z_0_fof}
\end{figure}

In Figure~\ref{fig:analytic_comp_z_0_fof} we show our universal FOF mass function relative to a number of fits available in the literature at $z=0$ , including the widely-used Press-Schechter (PS) \citep{1974ApJ...187..425P}, Sheth-Tormen (ST) \citep{2002MNRAS.329...61S} and Jenkins et al. \citep{2001MNRAS.321..372J} fits, as well as a number of more recent fits \citep{Warren:2005ey, Reed:2003sq, Reed:2006rw, Crocce:2009mg, Courtin:2010gx, Bhattacharya:2010wy, Angulo:2012ep}. We note that the fits by \citet{Reed:2006rw}, \citet{Crocce:2009mg} and \cite{Bhattacharya:2010wy} are redshift-dependent, while \citet{1974ApJ...187..425P, 2002MNRAS.329...61S} and \citet{Courtin:2010gx} are parameterised using $\delta_c(z)$. Aside from the classic, and less precise \cite{1974ApJ...187..425P} expression, our fit agrees to $\sim10\%$ with all the others for most of the mass range, and, for very massive clusters ($>10^{15}h^{-1}\mathrm{M}_{\odot}$) it matches very closely the recent \cite{Angulo:2012ep} fit based on the Millennium series of simulations. The effect of using data across all redshifts to create the universal fit is readily seen as the fit versus the $z=0$ data (shown in Figure~\ref{fig:fof_z_res}) is seen to be under-predicting mid-sized haloes. In fact, our $z=0$ data matches well a number of the predictions from the literature for these haloes. At high masses our data (Figure~\ref{fig:fof_z_res}) illustrates the large scatter from shot noise that is typically observed in the tail of the mass function.

With the exception of the PS, ST and \citet{Reed:2003sq} results, the largest discrepancy both with respect to our fit and overall scatter is for very large, rare haloes, whose statistics in most simulations is poor. A reliable result in this range requires very large simulation volumes of tens to hundreds of Gpc$^3$, comparable to the volume of the observable universe.

\begin{figure}
\includegraphics[width=3.3in]{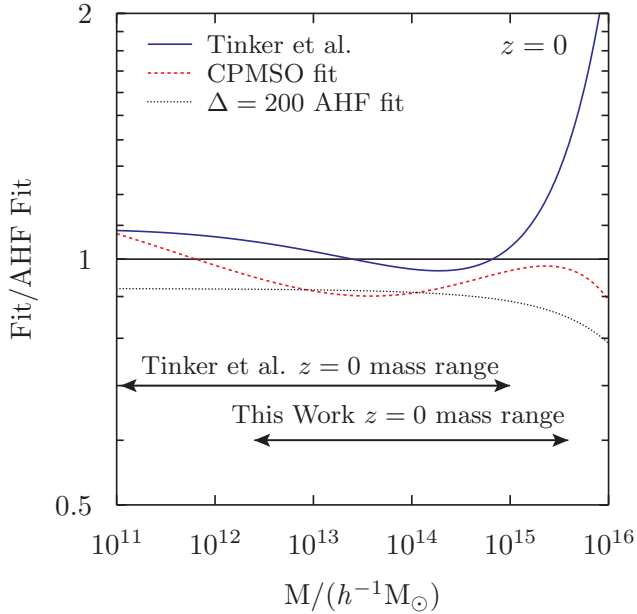}

\captionof{figure}{Ratios, at $z=0$, between the AHF $z = 0$ fit (solid black line) versus the \protect\cite{Tinker:2008ff} fit (solid blue line), the redshift parameterised CPMSO fit (dotted red line) and the AHF precise fit adjusted to $\Delta = 200$ using equation~\ref{Gamma_f:equ} (dotted black line).}
\label{fig:tinker_vs_watson}
\end{figure}

As we discussed above, the only other robust SO-based halo mass function fit is the one provided by \cite{Tinker:2008ff}. In Figure~\ref{fig:tinker_vs_watson} we show the ratio between our AHF $z=0$ fit and the \cite{Tinker:2008ff} fit at $z=0$. We also show how the CPMSO, redshift-dependent mass function compares to the AHF $z=0$ fit and the ratio between the AHF $z=0$ fit and the $\Delta = 200$ adjusted AHF $z=0$ fit (calculated using equation~\ref{Gamma_f:equ}). The \cite{Tinker:2008ff} fit is based on data in the range $-1.4<\mathrm{ln}\sigma^{-1}\le0.9$ for $z < 0.2$, while our fit includes data over the range $-0.55<\mathrm{ln}\sigma^{-1}\le1.35$ across all redshifts, or specifically $-0.55<\mathrm{ln}\sigma^{-1}\le1.05$ at $z=0$.

We show an agreement with \cite{Tinker:2008ff} to $\sim5\%$ for haloes in the mass range $10^{12}-10^{15}h^{-1}\mathrm{M}_{\odot}$. Past $\sim10^{15}h^{-1}\mathrm{M}_{\odot}$ we predict a lower collapsed mass fraction, although this is outside the range of the \cite{Tinker:2008ff} fitting data. The \cite{Tinker:2008ff} fit is based on $\Delta_{200}$ haloes and we show how our $\Delta _{200}$ prediction via equation~\ref{Gamma_f:equ}  compares to the underlying precision AHF plot. We note that this is slightly less congruent with the \cite{Tinker:2008ff} fit than the $\Delta _{178}$ prediction, although it is still in agreement to within 10\% for most of the mass range in question.

\subsection{Universality}

The results presented here are in line with other recent studies of the halo mass function that have addressed the question of universality, i.e. whether the mass function can be considered invariant over all redshifts or cosmological models when expressed in a suitable form. It is beyond the scope of this paper to discuss invariance under different cosmologies, for work on this see \cite{2001MNRAS.321..372J, Warren:2005ey, Tinker:2008ff} and in particular \cite{Courtin:2010gx} and \cite{Bhattacharya:2010wy}. The mass function has been found to depend only weakly on cosmology when couched in terms of FOF haloes with a fixed linking length. More study is required to address the question of how SO halo mass functions differ in varying cosmologies.

\begin{figure}
\includegraphics[width=3.3in]{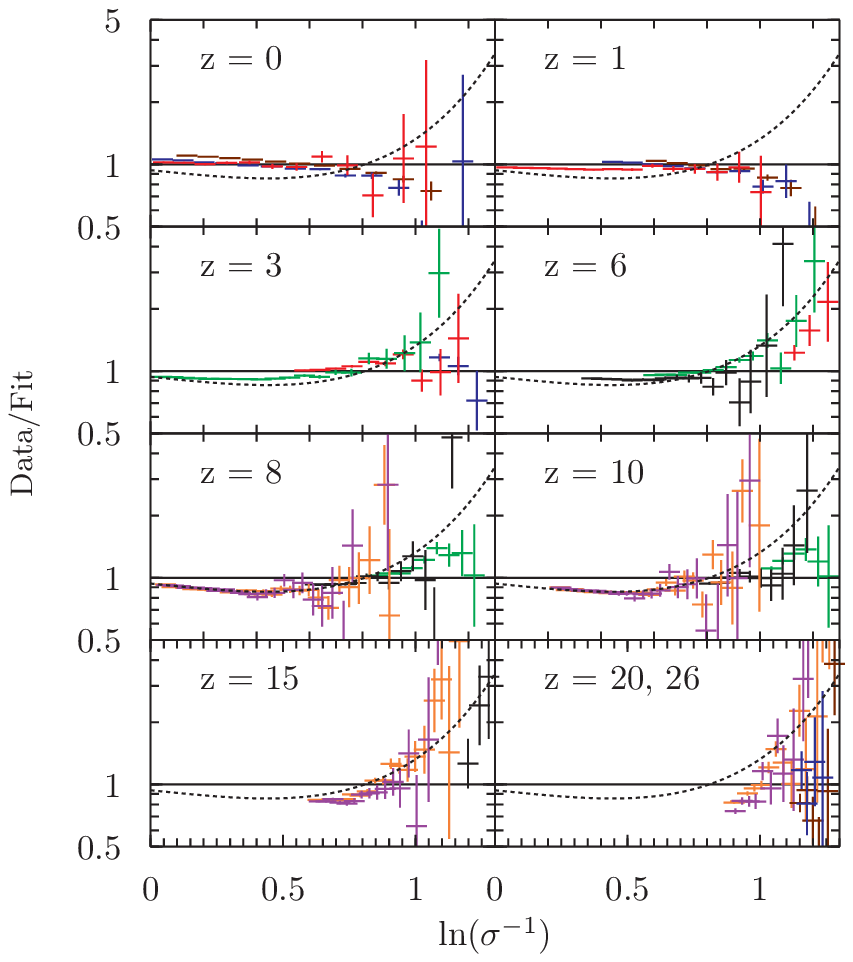}

\captionof{figure}{The redshift evolution of the FOF mass function versus the universal fit. The fit of \protect\cite{2002MNRAS.329...61S} is shown as a dashed line for reference.}
\label{fig:fof_z_res}
\end{figure}

We find that the assumption of halo mass function universality across redshift is approximately valid under certain conditions, but is violated in general, in qualitative agreement with the conclusions of other studies. \cite{2002ApJS..143..241W} highlighted that non-universality existed for most halo mass estimators; \cite{Tinker:2008ff} present an SO mass function that includes redshift-dependent parameters for $z \le 2.5$; similarly \cite{Crocce:2009mg} and \cite{Bhattacharya:2010wy} present FOF mass functions with redshift-dependent parameters for $z \le 2$; \cite{Reed:2006rw} produce a high-redshift mass function that contains a redshift-dependence via the effective slope of the power spectrum at the scale of the halo radius. We observe here that the FOF halo mass function for fixed linking length is close to universal for a very large redshift range (from $z=26$ to the present). \cite{2001MNRAS.321..372J} noted that taking a constant linking length keeps the mass function closer to a universal form, a result corroborated by \cite{Lukic:2007fc} who found that even for high redshifts the mass function of \cite{Warren:2005ey} was suitable for FOF haloes, despite it being calibrated on $z=0$ data. While we derived an universal fit for the FOF halo mass function for fixed linking length of 0.2, there clearly is some, albeit modest, redshift evolution about this fit. We show in Figure~\ref{fig:fof_z_res} the residuals between our FOF universal fit and our data for nine different redshifts. The amplitude of the data relative to the fit drops away markedly for lower ln$\sigma^{-1}$ values, to the extent that by $z\sim8$ and above the data is around 20\% lower than the prediction around ln$\sigma^{-1}\sim 0.5$.

In contrast, the SO-based halo mass functions are clearly not universal over the entire redshift range we consider, which was also previously noted by \cite{Tinker:2008ff} based on data over a smaller range in redshift and $\sigma$ than our data. Here we presented SO halo mass function fits that are applicable across a large range of redshifts, by explicitly introducing an $\Omega_m$- and redshift-dependent parameterisation of the mass function.

\subsection{Halofinder Comparison}
\label{sect:halofindercomp}

\subsubsection{Relationship Between SO and FOF Haloes}
\label{finder_comparison:sect}

The differences between an FOF halo and an SO halo, and the respective mass functions derived from haloes found using the two algorithms, are significant. A typical FOF halo is arbitrarily shaped and has been described in the literature as a demarkation of an isodensity contour in real space, with a value of $\Delta_{FOF}$ -- the overdensity of matter at the halo boundary -- that is under some debate. Estimates for $\Delta_{FOF}$ include $\Delta_{FOF} \approx 2b^{-3}$, i.e $\Delta_{FOF} \approx 250$ for the standard choice of $b=0.2$, \citep{1997ApJ...490..493N}; $\Delta_{FOF} \approx 0.48b^{-3}$, i.e. $\Delta_{FOF} \approx 60$ \citep{1994MNRAS.271..676L,Summers:1995ik,Audit:1998fj}; $\Delta_{FOF} \approx 74$ \citep{Warren:2005ey}; and $\Delta_{FOF} \approx 81.62$ \cite{More:2011dc}, all for $b=0.2$. In contrast the SO algorithm produces haloes that are by definition spherical and that adhere to a strict overdensity criterion, e.g. $\Delta=178$ times the background matter density. This overdensity refers to the average density of matter contained within the halo. Haloes are generally very centrally-concentrated, thus the overdensity of matter at the boundary of an SO halo is much lower than the mean overdensity and the relation between the two is dependent on the profile of the halo in question. \cite{More:2011dc} note that for a singular isothermal sphere (SIS) density profile, $\rho(r) \propto r^{-2}$, an overdensity at a halo's boundary of $\Delta_{edge} \sim 60$ corresponds to an enclosed overdensity of $\Delta_{halo} \sim 180$. This loosely links the value of $b$ in the FOF algorithm to $\Delta$ in the SO algorithm --by assuming that the FOF halo is spherical-- but the simplified SIS profile does not reproduce simulated dark matter haloes well. More appropriate profiles include the TIS profile \citep{1999MNRAS.307..203S,2001MNRAS.325..468I}, the Einasto profile \citep{2004MNRAS.349.1039N,2008MNRAS.387..536G,2010MNRAS.402...21N} and the NFW profile. \cite{More:2011dc} used the latter, sampling NFW haloes with random particle realisations and show that their value of $\Delta_{FOF} \approx 81.62$ corresponds to a range of $\Delta_{halo}$ between $\approx 200-600$ depending on the concentration parameter of the NFW haloes. \cite{Lukic:2008ds} use a similar approach and were able to recover, to 5\% accuracy, an SO mass function from FOF haloes by individually relating the FOF halo masses to their SO counterparts. The translation between the two was based on the concentration parameter of the haloes' NFW profiles and three empirically fitted parameters. 

\subsubsection{Choice of Linking Length and Overdensity Criterion}
\label{LL_and_Delta:sect}

\begin{figure}
\includegraphics[width=3.3in]{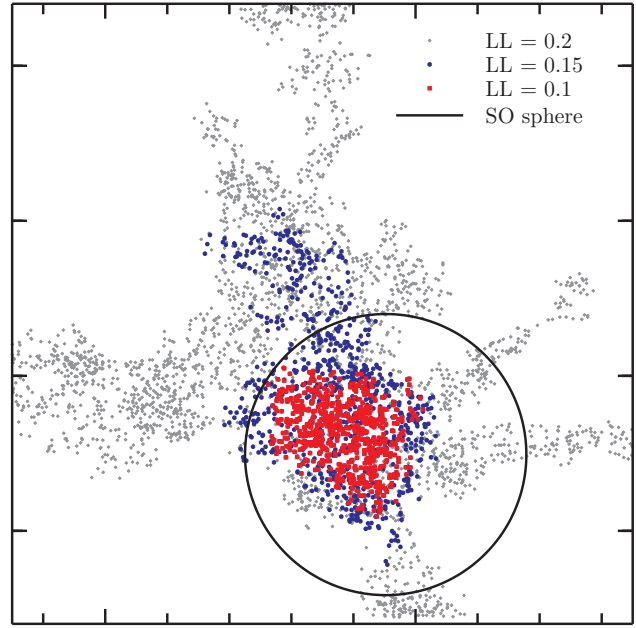}
\captionof{figure}{Image of a large halo in the 20 $h^{-1}$Mpc box at $z=8$. The circle represents the extent of the $\Delta_{178}$ cutoff used in CPMSO. The z direction has been projected onto the x-y plane. The CPMSO halo mass is 3.1$\times$10$^{10}h^{-1}\mathrm{M}_{\odot}$ and it contains 9.3 million particles. The dots represent aggregations of at least 20 particles found in the FOF version of the same halo. Grey shows the halo captured with a linking length of 0.2, blue 0.15 and red 0.1. The masses (particle counts) are 4.8$\times$10$^{10}h^{-1}\mathrm{M}_{\odot}$ (13.1 million), 3.7$\times10^{10}h^{-1}\mathrm{M}_{\odot}$ (8.9 million) and 2.1$\times10^{10}h^{-1}\mathrm{M}_{\odot}$ (5.8 million) for $b=0.2, 0.15, 0.1$ respectively.}
\label{fig:linklength_pic}
\end{figure}

Given these analyses it is not surprising to observe a marked difference in the physical structure of FOF and SO haloes. Figure~\ref{fig:linklength_pic} shows a large halo from our 20 $h^{-1}$Mpc simulation at $z=8$ captured by the CPMSO and the FOF halo finders, respectively. We show the FOF halo found with linking lengths of 0.2, 0.15, and 0.1 in grey, blue and red respectively. This is the second largest halo in our volume and it illustrates particularly well the differences between the two algorithms. For this halo at this redshift a linking length of 0.2 is far too aggressive and significant overlinking has occurred.

\begin{figure*}
\includegraphics[width=6.6in]{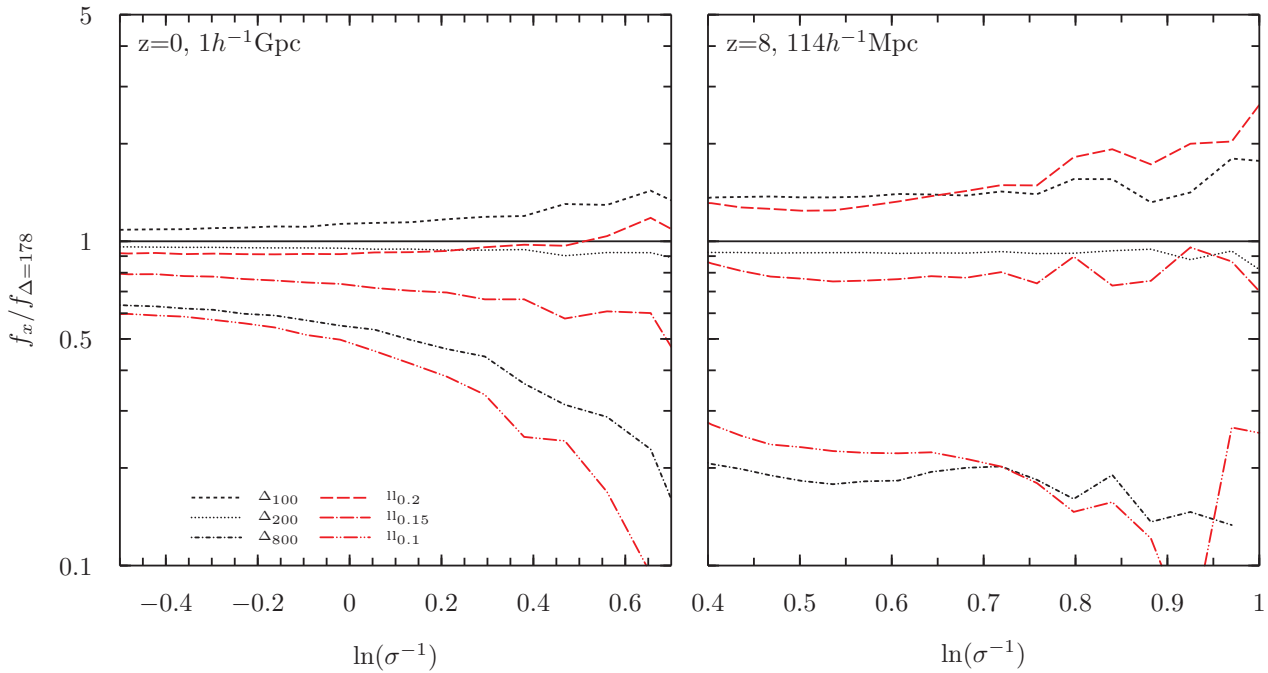}
\captionof{figure}{Comparison of mass functions from the 1$h^{-1}$Gpc simulation at $z=0$ and the 114$h^{-1}$Mpc simulation at $z = 8$ using AHF with a variable overdensity and FOF with a variable linking length. All ratios are plotted on a base of the $\Delta = 178$ results.}
\label{fig:ll_vs_delta}
\end{figure*}

In Figure~\ref{fig:ll_vs_delta} we compare the halo mass functions based on FOF haloes with linking lengths of 0.2, 0.15, 0.1 and SO haloes with overdensity choices of 100, 178, 200 and 800, at $z=0$ from the 1$h^{-1}$Gpc box (left-hand panel) and at $z=8$ from the 114$h^{-1}$Mpc box (right-hand panel). The $z=0$ result shows that, for a suitable range of $\sigma$, an overdensity of 178 is comparable to a linking length of 0.2 to within 10\%. However, for a 0.2 linking lengh there appears to be a trend towards lower overdensities for the higher mass haloes. This is likely due to the increasing influence of overlinking on the masses of the larger FOF haloes. The $z=8$ results show that a linking length of 0.2 is not at all complimentary to an overdensity choice of 178 in the EdS regime of structure growth. In fact, we see that an overdensity of 100 is more consistent with a linking length of 0.2 at $z=8$. The equivalent linking length for an overdensity of 178 lies between 0.2 and 0.15 and much closer to 0.15 than to 0.2, and a linking length of 0.1 roughly corresponds to overdensity of 800. These results are consistent the with result of \cite{2008MNRAS.385.2025C}, namely that, at $z = 10$, the mass functions of FOF and SO haloes are similar when an overdensity choice of $\Delta = 180$ and a linking length of $ll=0.168$ are used. Given these results we expect there to be an evolution in the relationship between overdensity and linking length. The empirical results of \cite{Lukic:2008ds}, \cite{Courtin:2010gx} and \cite{More:2011dc} all contain a redshift dependence (\cite{Courtin:2010gx} via $\Delta(z)$, and the others via the concentration parameter, $c(z)$). Whilst the \cite{More:2011dc} study looked at haloes up to $z=2.5$ further study needs to be undertaken to investigate the suitability of existing relations between linking length and overdensity at higher redshifts.

\subsubsection{SO vs. FOF Mass Functions}

The differences we observed above between haloes found using the two approaches filter through to secondary results derived from halo catalogues, including the mass function. Figure~\ref{fig:fofcor_vs_ahf} shows the systematic difference between the AHF ($\Delta = 178$) and the FOF ($b=0.2$) mass functions over a wide range of redshifts. There is a close similarity between the two mass functions at $z = 0$, especially for lower mass haloes, with the collapsed fraction of mass in higher-mass haloes greater in the FOF case. Gradually, as we move to higher redshifts, we see the amplitude of the SO mass function fall to around 80\% of the FOF mass function at lower masses, with a much more marked decrease at higher masses. Past redshifts of around $z\sim6$ the high mass tail of the SO mass function is around 50\% lower than that of the FOF mass function. The causes of the difference between the two mass functions can be summarised as contributions from (1) the relationship between the masses of a given SO halo and its FOF counterpart, which depend on the choices of $b$ and $\Delta$ and also on the concentration parameter of the haloes; (2) the amount of overlinking of FOF haloes; (3) the relative mass difference in the two halo types that arises from pseudo mass evolution, as discussed in \S~\ref{hm_redshift_dep}; and (4) other systematic effects including the SO algorithm not correctly interpreting the properties of non-spherical haloes and both algorithms failing to reliably describe merging systems. \cite{2001A&A...367...27W}, \cite{2002ApJS..143..241W}, \cite{Lukic:2008ds} and \cite{More:2011dc} have addressed the first factor in detail. \cite{2012MNRAS.423.3018P} provide a detailed exposition of the evolution of the concentration parameter over redshift. Their findings illustrate that halo concentrations lie on a characteristic `U' shape in the c-ln$\sigma^{-1}$ plane. This shape exhibits modest evolution in redshift, with the concentration of the minimum becoming slightly smaller at higher redshifts. \cite{1985ApJ...292..371D,1995ApJ...455....7M,1996MNRAS.281..716C,Lukic:2008ds} have investigated the second and, to some extent, the fourth factors. The third factor has not been studied in detail and remains a topic for future study.
 
 \section{Acknowledgements}
 
We thank Volker Springel for his permission to use the FOF halofinder from the Gadget-3 code and Stefan Gottl\"ober, Leonidas Christodoulou and Jos\'e Maria Diego for helpful feedback and discussion. We thank Aurel Schneider for meticulously checking the manuscript, the anonymous referee for very helpful comments and Steffen Knollman for his help in applying the AHF halofinder to our data. Also we are very grateful and thank in particular Sergio Palomares-Ruiz for his feedback on the earlier drafts of the paper. The Jubilee simulation has been performed and analysed at the Neumann Institute for Computing (NIC) J\"ulich (Germany). All other simulations were undertaken at the Texas Advanced Computing Center (TACC) at The University of Texas at Austin under TeraGrid grants. WW thanks The Southeast Physics Network (SEPNet) for providing funding for his research. ITI was supported by The SEPNet and the Science and Technology Facilities Council grants ST/F002858/1 and ST/I000976/1. AK is supported by the Spanish Ministerio de Ciencia e Innovacion (MICINN) in Spain through the Ramon y Cajal programme as well as the grants AYA2009-13875-C03-02, AYA2009-12792-C03-03, CSD2009-00064, and CAM S2009/ESP-1496. GY acknowledges support from MINECO (Spain) under research grants AYA2009-13875-C03-02, FPA2009-08958 and Consolider Ingenio SyeC CSD2007-0050. 

\bibliographystyle{mn}

\end{document}